\documentclass[12pt]{article}
\usepackage{graphicx,epsf}
\usepackage{lscape}
\usepackage{citesort}

\begin{document}

\def\ds{\displaystyle}
\def\beq{\begin{equation}}
\def\eeq{\end{equation}}
\def\bea{\begin{eqnarray}}
\def\eea{\end{eqnarray}}
\def\beeq{\begin{eqnarray}}
\def\eeeq{\end{eqnarray}}
\def\ve{\vert}
\def\vel{\left|}
\def\ver{\right|}
\def\nnb{\nonumber}
\def\ga{\left(}
\def\dr{\right)}
\def\aga{\left\{}
\def\adr{\right\}}
\def\lla{\left<}
\def\rra{\right>}
\def\rar{\rightarrow}
\def\nnb{\nonumber}
\def\la{\langle}
\def\ra{\rangle}
\def\ba{\begin{array}}
\def\ea{\end{array}}
\def\tr{\mbox{Tr}}
\def\ssp{{\Sigma^{*+}}}
\def\sso{{\Sigma^{*0}}}
\def\ssm{{\Sigma^{*-}}}
\def\xis0{{\Xi^{*0}}}
\def\xism{{\Xi^{*-}}}
\def\qs{\la \bar s s \ra}
\def\qu{\la \bar u u \ra}
\def\qd{\la \bar d d \ra}
\def\qq{\la \bar q q \ra}
\def\gGgG{\la g^2 G^2 \ra}
\def\q{\gamma_5 \not\!q}
\def\x{\gamma_5 \not\!x}
\def\g5{\gamma_5}
\def\sb{S_Q^{cf}}
\def\sd{S_d^{be}}
\def\su{S_u^{ad}}
\def\ss{S_s^{??}}
\def\sbp{{S}_Q^{'cf}}
\def\sdp{{S}_d^{'be}}
\def\sup{{S}_u^{'ad}}
\def\ssp{{S}_s^{'??}}
\def\sig{\sigma_{\mu \nu} \gamma_5 p^\mu q^\nu}
\def\fo{f_0(\frac{s_0}{M^2})}
\def\ffi{f_1(\frac{s_0}{M^2})}
\def\fii{f_2(\frac{s_0}{M^2})}
\def\O{{\cal O}}
\def\sl{{\Sigma^0 \Lambda}}
\def\es{\!\!\! &=& \!\!\!}
\def\ap{\!\!\! &\approx& \!\!\!}
\def\ar{&+& \!\!\!}
\def\ek{&-& \!\!\!}
\def\kek{\!\!\!&-& \!\!\!}
\def\cp{&\times& \!\!\!}
\def\se{\!\!\! &\simeq& \!\!\!}
\def\eqv{&\equiv& \!\!\!}
\def\kpm{&\pm& \!\!\!}
\def\kmp{&\mp& \!\!\!}


\def\simlt{\stackrel{<}{{}_\sim}}
\def\simgt{\stackrel{>}{{}_\sim}}


\title{
         {\Large
                 {\bf
Analysis of the rare semileptonic  $B_c \rar P(D,D_s)
l^{+}l^{-}/\nu\bar{\nu}$ decays within  QCD sum rules
                 }
         }
      }

\author{\vspace{1cm}\\
{\small K. Azizi \thanks {e-mail: e146342 @ metu.edu.tr}~\,} \\
{\small  Department of Physics, Middle East Technical University,
06531 Ankara, Turkey}\\
{\small R. Khosravi \thanks {e-mail: khosravi.reza @ gmail.com}~\,}\\
 {\small Physics Department , Shiraz University, Shiraz 71454,
Iran}\\}
\date{}

\begin{titlepage}
\maketitle \thispagestyle{empty}

\begin{abstract}
Considering the gluon condensate corrections, the form factors
relevant to the semileptonic rare $B_c \rar D,D_s(J^{P}=0^{-})
l^{+}l^{-}$ with $l=\tau,\mu,e$ and $B_c \rar
D,D_s(J^{P}=0^{-})\nu\bar{\nu}$ transitions are calculated in the
framework of the three point QCD sum rules. The heavy quark
effective theory limit of the form factors are computed. The
branching fraction of these decays are also evaluated and compared
with the predictions of the relativistic constituent quark model.
Analyzing of such type transitions could give useful information
about the strong interactions inside the pseudoscalar $D_{s}$ meson
and its structure.
\end{abstract}

\end{titlepage}

\section{Introduction}
With the chances that in the future a large amount of Bc mesons will
be produced at LHC (with the luminosity values of ${\cal
L}=10^{34}cm^{-2}s^{-1}$ and $\sqrt{s}=14\rm TeV$, the number of
$B_c^{\pm}$ mesons is expected to be about $10^{8}\sim10^{10}$ per
year \cite{Du,Stone}), one might explore the rare Bc decays to
pseudoscalar   $(D,D_s) $ and $l^{+}l^{-}/\nu\bar{\nu}$. Such types
transitions could be useful because of the following reasons: 1)
Analyzing of such type transitions could give valuable information
about the nature of the pseudoscalar $D_{s}$ meson and the strong
interactions inside it. 2) The form factors of these transitions
could be used in the study of the polarization asymmetries, CP and T
violations. 3) These will provide a new framework for more precise
calculation of the Cabibbo-Kobayashi-Maskawa (CKM) matrix elements
$V_{tq}$ (q = d, s, b) and  leptonic decay constants of $D_{s,d}$
and $B_c$ mesons. 4) These transitions occur at loop level in
standard model (SM) via the flavour changing neutral current (FCNC)
transitions of $b \rightarrow s, d$, which are sensitive to the new
physics beyond the SM, so these  decays are useful to constrain the
parameters beyond the SM. 5) A possible forth generation, SUSY
particles \cite{Buchalla} and light dark matter \cite{Bird} might
contribute to the loop transitions of $b \rightarrow s, d$.

The $B_{c}$, is the only meson containing two heavy quarks with
different charge and flavours  and it is the lowest bound state of b
and  c quarks, so  its decay modes properties are expected to be
different than  flavour neutral mesons. Since the excited levels of
$\bar b c$ lie below the threshold of decay into the pair of heavy B
and D mesons, such states decay weakly and they have no annihilation
decay modes due to the electromagnetic and strong interactions (for
more about the physics of the $B_{c}$ meson see for example
\cite{Gershtein}). This paper describes the annihilation of the
$B_{c}$ into the pseudoscalar $(D,D_s)l^{+}l^{-}/\nu\bar{\nu} $ in
the framework of the three point QCD sum rules as a non-perturbative
approach based on the fundamental QCD Lagrangian. This transitions
are parameterized in terms of some form factors calculation of which
plays crucial role in the analyzing of those decay channels. These
decays at quark level proceed by the loop $b \rightarrow s, d$ in
the SM with the intermediate u, c and t quarks and the main
contribution comes from the intermediate top quark. These decay
modes have also been studied in the relativistic constituent quark
 model (RCQM)
\cite{tt1}. Some other possible channels such as  $B_{c}\rightarrow
l \overline{\nu}\gamma$, $B_{c}\rightarrow \rho^{+}\gamma$,
$B_{c}\rightarrow K^{\ast+}\gamma$, $B_{c}\rightarrow
B_{u}^{\ast}l^{+}l^{-}$, $B_{c}\rightarrow B_{u}^{\ast}\gamma $,
$B_{c}\rightarrow D_{s,d}^{\ast}\gamma $, $B_{c}\rightarrow
D_{s,d}^{\ast}l^{+}l^{-} $ and $B_{c}\rightarrow X\nu\bar{\nu} $
with $X$ be  axial vector particle, $D_{s1}(2460)$, and vector
particles, $D^*,D^*_s$ are studied in the light cone or traditional
QCD sum rules methods in
\cite{Aliev1,Aliev2,Aliev3,Alievsp,azizi1,azizi2,azizi3},
respectively. For a  set of exclusive nonleptonic and semileptonic
decays of the $B_{c}$ meson, which have been studied in the
relativistic constituent quark model see \cite{Ivanov}.

The content of paper is as follows: In section 2, we calculate the
sum rules for the related form factors considering the gluon
correction contributions to the corrolation function. The light
quark condensate contributions are killed applying the double borel
transformations with respect to momentum of the initial and final
states. The heavy quark effective theory (HQET) limit of the form
factors are presented in section 3. Section 4 depicts our numerical
analysis of the form factors and their comparison with the HQET
limit of them, results, discussions and comparison of our results
with the prediction of the RCQM model.
\begin{figure}
\vspace*{-1cm}
\begin{center}
\includegraphics[width=11cm]{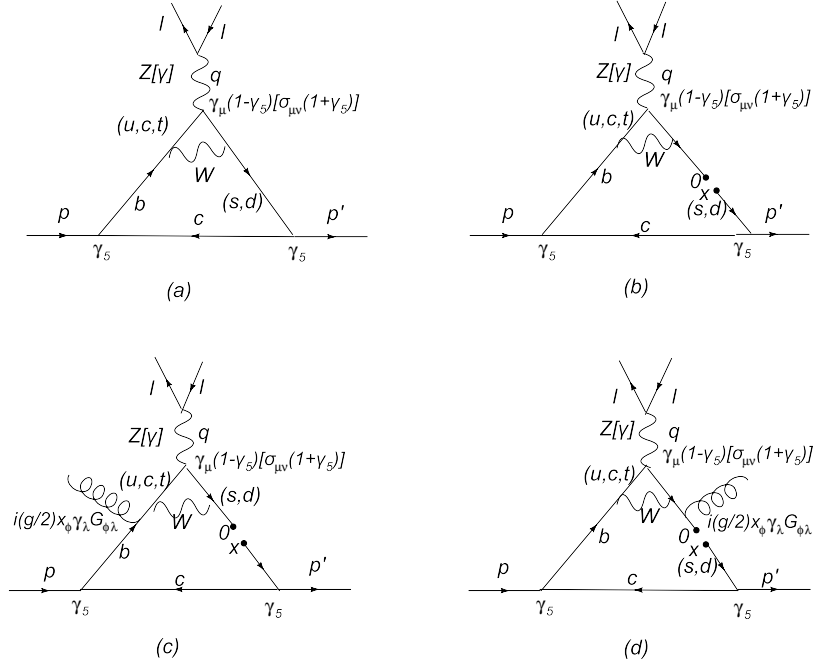}
\end{center}
\caption{loop diagrams for $B_c \rar (D,D_s)l^{+}l^{-}/
\nu\bar{\nu}$ transitions, bare loop (diagram a) and light quark
condensates (without any gluon diagram b and with one gluon emission
diagrams c, d)} \label{fig1}
\end{figure}

\section{QCD Sum rules for transition form factors of the $B_c \rar (D,D_s)l^{+}l^{-}/\nu\bar{\nu} $ }

At quark level, the processes $B \rightarrow P l^{+}l^{-}/\nu\bar{\nu}$$%
(P=D,D_s)$ are described  by the loop  $b \rightarrow q_{_i}
l^{+}l^{-}/\nu\bar{\nu}$ transitions, ($q_{_1}=d~,~q_{_2}=s$) in the
SM (see Fig.1), and receive contributions from photon and
$Z$-penguin and box diagrams for $l^{+}l^{-}$  and only $Z$-penguin
and box diagrams for $\nu\bar{\nu}$. These loop transitions occur
via the intermediate u, c, t quarks , where dominant contribution
comes from intermediate top quark. The effective Hamiltonian
responsible for $b \rightarrow q_{_i} l^+
l^-$ decays is described in terms of the Wilson coefficients,  $%
C^{eff}_{7},C^{eff}_{9}$ and $C_{10} $ as

\begin{eqnarray}  \label{e1}
\mathcal{H}_{eff} &=& \frac{G_{F}\alpha}{2\sqrt{2} \pi} V_{tb}V_{tq_{_i}}^\ast %
\Bigg[ C_9^{eff} \, \bar q_{_i} \gamma_\mu (1-\gamma_5) b \, \bar
\ell \gamma_\mu \ell + C_{10}~ \bar q_{_i} \gamma_\mu (1-\gamma_5)
b \, \bar \ell \gamma_\mu \gamma_5
\ell  \nonumber \\
&-& 2 C_7^{eff} \frac{m_b}{q^2}~ \bar q_{_i} ~i\sigma_{\mu\nu}
q^\nu (1+\gamma_5) b \, \bar \ell \gamma_\mu \ell \Bigg]~,
\end{eqnarray}
where $G_{F}$ is the Fermi constant, $\alpha$ is the fine structure
constant at  $Z$ mass scale, and $V_{ij}$ are elements of the CKM
matrix. For $\nu\bar{\nu}$ case, only term containing $C_{10}$ is
considered. The amplitudes for  for $B_c \rightarrow P
l^{+}l^{-}/\nu\bar{\nu}$  decays are obtained by  sandwiching of Eq.
(\ref{e1}) between initial and final meson states:

\begin{eqnarray}\label{e2}
\mathcal{M}&=&\frac{G_{F}\alpha }{2\sqrt{2}\pi
}V_{tb}V_{tq_{_i}}^{\ast } \Bigg[C_{9}^{eff}\,<P(p')\mid
~\bar{q_{_i}}\gamma _{\mu }(1-\gamma _{5})b\mid B_{c}(p)>\bar{
\ell}\gamma _{\mu }\ell\nonumber
\\\nonumber &&+ ~C_{10}~<P(p')\mid \bar{q_{_i}}\gamma _{\mu }(1-\gamma _{5})b\mid
B_{c}(p)>\bar{\ell} \gamma _{\mu }\gamma _{5}\ell
\\ &&-2~C_{7}^{eff}~\frac{m_{b}}{q^{2}}<P(p')\mid \bar{q_{_i}}~i\sigma _{\mu \nu
}q^{\nu }(1+\gamma _{5})b\mid B_{c}(p)>\bar{\ell}\gamma _{\mu }\ell
\Bigg].
\end{eqnarray}
Next, we calculate the the matrix elements $<P(p')\mid \bar{q_{_i}%
}\gamma _{\mu }(1-\gamma _{5})b\mid B_{c}(p)>$ and $<P(p')\mid
\bar{q_{_i}}~i\sigma _{\mu \nu }q^{\nu }(1+\gamma _{5})b\mid
B_{c}(p)>$ appearing in above equation. The parts of transition
currents containing $\gamma _{5}$ don't contribute, so we consider
only  $~\bar{q_{_i}}~\gamma _{\mu }b~$ and also $\bar{q_{_i}}\sigma
_{\mu \nu }q^{\nu }b$ parts. Considering  Lorentz and parity
invariances, this matrix elements can be parameterized in terms of
the form factors as:
\begin{equation}  \label{e3}
<P(p^{\prime}){\mid }\bar{q_{_i}}{\gamma }_{\mu }b{\mid }%
B_{c}(p)>=-(~\mathcal{P}_{\mu }f_{+}(q^{2})+~{q}_{\mu
}f_{-}(q^{2})),
\end{equation}
\begin{equation}\label{e4}
<P(p^{\prime})\,|\,\bar{q_{_i}}\,i\sigma _{\mu \nu }q^{\nu
}\,b\,|\,B_{c}(p)>= \, \frac{f_T(q^{2})}{m_B+m_P} \Big[
\mathcal{P}_\mu q^2 - q_\mu (m_{B_c}^2-m_P^2) \Big],
\end{equation}
where $f_{+}(q^{2}) , f_{-}(q^{2})$ and $f_{T}(q^{2})$ are the transition form factors, $%
\mathcal{P}_{\mu }=(p+p^{\prime })_{\mu }$ and $q_{\mu
}=(p-p^{\prime
})_{\mu }$. Here, we should mention that for $\nu\bar{\nu}$ case the form factor $f_{T}(q^{2})$
does'nt contribute since it is related to the photon vertex ($\sigma _{\mu \nu }q^{\nu }$).
To calculate  the form factors $f_{+}(q^{2})$, $f_{-}(q^{2})$ and $%
f_{T}(q^{2})$, we start with the following correlation function:

\begin{eqnarray}  \label{e5}
\Pi_{\mu}^{V,T} = i^2\int d^{4}xd^4ye^{-ipx}e^{ip^{\prime}y}\langle
0 \vert \mathcal{T} \left\{ J_{P }(y) J_\mu^{V,T}(0)
J^{\dag}_{B_c}(x) \right\} \vert 0 \rangle~,
\end{eqnarray}

where $J _{P }(y)=\bar{c}\gamma_{5} q_{_i}$ ($q_{_i}= s$~or~$d$) and
$J_{B_{c}}(x)= \bar{c}\gamma_{5}b$  are the interpolating currents
of the $P$ and $B_{c}$ messons and
$J_{\mu}^{V}=\bar{q_{_i}}\gamma_{\mu}b ~$ and $
J_{\mu}^{T}=\bar{q_{_i}}i\sigma _{\mu \nu }q^{\nu }b$ are transition
currents. From the general philosophy of the QCD sum rules, we can
calculate the above mentioned corrolator in two languages: 1) hadron
language called the physical or phenomenological side, 2) quark
gluon language which is the QCD or theoretical side. Equating two
sides and applying the double Borel transformations with respect to
the momentum of the initial and final states to suppress the
contribution of the higher states and continuum, we get sum rules
expressions for our form factors. The phenomenological part can be
obtained  by inserting the complete set of intermediate states with
the same quantum numbers as the currents $J_{P}$ and $J_{B_{c}}$. As
a result of this procedure
\begin{eqnarray}  \label{e7}
\Pi_\mu^{V,T}(p^2,p^{\prime 2},q^2) \!\!\! &=& \!\!\!
\frac{\displaystyle \left< 0 \left| J_{P} \right| P(p^{\prime})
\right> \left< P(p^{\prime}) \left| J_{\mu}^{V,T} \right| B_c(p)
\right> \left< B_c(p) \left| J^{\dag}_{B_c} \right| 0 \right>} {\displaystyle %
(m_{P}^2 - p^{\prime 2}) (m_{B_c}^2 - p^2)},
\end{eqnarray}
is obtained. The following matrix elements are defined in terms of
the leptonic decay constants of the $P$ and $B_c$ mesons as:
\begin{eqnarray}  \label{e9}
\left< 0 \left| J_{P} \right| P \right> \!\!\! &=&  \!\!\! -i\frac{%
\displaystyle f_{P} m_{P}^2}{\displaystyle m_{c}+m_{q_{i}}}~,
\nonumber \\
\left< 0 \left| J_{B_c} \right| B_c \right> \!\!\! &=& \!\!\! -i\frac{%
\displaystyle f_{B_c} m_{B_c}^2}{\displaystyle m_b+m_c}~.
\end{eqnarray}

 Using Eqs. (\ref{e3}), (\ref{e4})
and (\ref{e9}) in Eq. (\ref{e7}), we obtain
\begin{eqnarray}  \label{e10}
\Pi_\mu^{V} (p^2,p^{\prime 2},q^2) \!\!\! &=& \!\!\!
-\frac{\displaystyle f_{B_c}
m_{B_c}^2}{\displaystyle (m_b+m_c)(m_{c}+m_{q_{i}})} \frac{\displaystyle f_{P} m_{P}^{2}}{%
\displaystyle (m_{P}^2 - p^{\prime 2}) (m_{B_c}^2 - p^2)} \Big[ f_+ \mathcal{%
P}_\mu + f_- q_\mu \Big] + ...~,
\end{eqnarray}
\begin{eqnarray}  \label{e11}
\Pi_\mu^{T} (p^2,p^{\prime 2},q^2) \!\!\! &=& \!\!\!
\frac{\displaystyle f_{B_c}
m_{B_c}^2}{\displaystyle (m_b+m_c)(m_{c}+m_{q_{i}})} \frac{\displaystyle f_{P} m_{P}^{2}}{%
\displaystyle (m_{P}^2 - p^{\prime 2}) (m_{B_c}^2 - p^2)} \nonumber\\&\times&\Big[ \frac{ f_{T}}{(m_{B_c} + m_{P})} ~[q^2\mathcal{%
P}_\mu -(m_{B_c}^2 - m_{P}^2)q_{\mu}]\Big] + ...~.
\end{eqnarray}
For extracting the expressions for form factors $f_{+}(q^{2})$ and
$f_{-}(q^{2})$, we choose the coefficients of the structures
$\mathcal{P}_{\mu }$  and $q_\mu$ from $\Pi_\mu^{V} (p^2,p^{\prime
2},q^2)$, respectively and the structure $q_\mu$ from $\Pi_\mu^{T}
(p^2,p^{\prime 2},q^2)$ is considered for the form factor
$f_{T}(q^{2})$. Therefore, the correlation functions are written in
terms of the selected structures as: \bea \label{e7309}
\Pi_{\mu}^{V}(p^2,p'^2,q^2) \es \Pi_+\mathcal{P}_{\mu }  + \Pi_{-}
q_\mu  +...~,\eea \bea \label{e7309} \Pi_{\mu}^{T}(p^2,p'^2,q^2) \es
\Pi_T q_\mu  +...~.\eea

On the other side, to calculate the QCD part of correlation
function, we evaluate the three--point correlator by the help of the
operator product expansion (OPE) in the deep Euclidean region, where
$p^2 \ll (m_b + m_c)^2$ and $p'^2 \ll (m_c+m_{q_{i}})^2$. For this
aim, we write each $\Pi_{i}$ function
 in terms of the perturbative and non-perturbative parts as:

\bea \label{e7316} \Pi_{i}(p^2,p'^2,q^2) =
\Pi_{i}^{per}(p_1^2,p_2^2,q^2) +\Pi_{i}^{nonper}(p^2,p'^2,q^2)~,
\eea where  $i$ stands for $+$, $-$ and $T$ and non-perturbative
part contains  the light quark ($<\bar qq>$) and gluon ($<G^2>$)
condensates. For the perturbative part, the bare loop diagram (Fig.
1 a) is considered, however, diagrams b, c, d in Fig. 1 are
correspond to the light quark condensates contributing to the
correlator.  In principle, the light quark condensate diagrams give
contributions to the correlation function, but applying double Borel
transformations omits their contributions, hence as first
non-perturbative correction, we   consider the gluon condensate
diagrams (see Fig. 2 a, b, c, d, e, f).

By the help of  the double dispersion representation, the bare--loop
contribution is written as \bea \label{e7317} \Pi_{i}^{per} = -
\frac{1}{(2 \pi)^2} \int ds^\prime \int ds \frac{\rho_{i}^{per}
(s,s^\prime, Q^2)}{(s-p^2) (s^\prime - p^{\prime 2})}   + \mbox{\rm
subtraction terms}~, \eea  where $Q^2=-q^2$. The spectral densities
$\rho_{i}^{per} (s,s^\prime,Q^2)$ are calculated by  the help of the
Gutkovsky rule, i.e.,  the propagators are replaced by Dirac--delta
functions \bea \label{e7319} \frac{1}{p^2-m^2} \rar -2i\pi
\delta(p^2-m^2)~,\eea expressing  that all quarks are real. The
integration region in Eq. (\ref{e7317}) is obtained by requiring
that the argument of three delta vanish, simultaneously. This
condition results in   the following inequality \bea \label{e7318}
-1 \le \frac{2 s s^\prime + (s + s^\prime + Q^2 )(m_b^2 - m_c^2 - s)
+ 2 s (m_c^2-m^2_{q_{i}})}{\lambda^{1/2} (s,s^\prime,-Q^2)
\lambda^{1/2}(m_b^2,m_c^2,s)} \le +1~, \eea where
$\lambda(a,b,c)=a^2+b^2+c^2-2ab-2ac-2bc$. From this inequality, to
use in the lower and upper limit of the integration over $s$ in
subtractions, it is easy to express $s$ in terms of $s'$ i.e.
$f_{\pm}(s')$ in the $s-s'$ plane.

  Straightforward calculations end up in  the following results   for the
spectral densities:
\begin{eqnarray*}
\rho _{+}^{V}(s,s^{\prime },q^{2})\!\!\!
&=&\!\!\!\mathit{I_{0}}\,{Nc}\,\{\Delta
+\Delta ^{\prime }+ \\
&&-2\,{m_{c}}\,[(+2+{E_{1}}+{E_{2}}){m_c}-(1+{E_{1}}+{E_{2}}){m_{q_{_i}}}] \\
&&+2\,{m_{b}[}(1+{E_{1}}+{E_{2}}){m_{c}-({E_{1}}+{E_{2}})m_{q_{_i}}]} \\
&&+({E_{1}}+{E_{2}})u\},
\end{eqnarray*}
\begin{eqnarray*}
\rho _{-}^{V}(s,s^{\prime },q^{2})\!\!\! &=&\!\!\!\mathit{I_{0}}\,{Nc}\,\{-{%
\Delta }+{\Delta ^{\prime }} \\
&&-2\,{m_{c}}\,[({E_{2}}-{E_{1}-1}){m_{q_{_i}}}+({E_{1}}\,-{E_{2}}\,){m_{c}]}
\\
&&-2{m_{b}[(1-E_{1}}\,+{E_{2}}\,){m_{c}+}({E_{1}}\,-{E_{2}}\,){m_{q_{_i}}]} \\
&&+({E_{1}}-{E_{2}})u\},
\end{eqnarray*}
\begin{eqnarray*}
\rho _{T}^{T}(s,s^{\prime },q^{2}) &=&-{I_{0}}\,{N_{c}}\,\{{\Delta }\,({%
2m_{c}-m_{b}}-{m_{q_{_i}}})+{\Delta ^{\prime }}\,(m_{b}-2m_{c}+{m_{q_{_i}}}) \\
&&+2[\,{m_{c}}({E_{1}}-{E_{2}}-1)\,+\,{m_{q_{_i}}}\,({E_{2}}-{E_{1}})]\,s \\
&&-2\,[{m_{b}}({E_{1}}-{E_{2}})-{m_{c}}({E_{1}}-{E_{2}}+1)]{s^{\prime }} \\
&&+({E_{1}}-{E_{2}})({m_{b}}-2\,{m_{c}}+{m_{q_{_i}}})u\},
\end{eqnarray*}
where
\begin{eqnarray}  \label{12}
I_{0}(s,s^{\prime},Q^2)&=&\frac{1}{4\lambda^{1/2}(s,s^{\prime},Q^2)}
,
\nonumber \\
\lambda(s,s^{\prime},Q^2)&=&s^2+s'^{2}+Q^4+2sQ^2+2s^{\prime}Q^2-2ss^{%
\prime},  \nonumber \\
E_{1}&=&\frac{1}{\lambda(s,s^{\prime},Q^2)}[2s^{\prime}\Delta-\Delta^{%
\prime}u] ,  \nonumber \\
E_{2}&=&\frac{1}{\lambda(s,s^{\prime},Q^2)}[2s\Delta^{\prime}-\Delta
u] ,
\nonumber \\
u&=&s+s^\prime+Q^2 ,  \nonumber \\
\Delta&=&s+m_c^2-m_b^2 ,  \nonumber \\
\Delta^{\prime}&=&s^\prime+m_c^2-m^2_{q_{_i}} , \nonumber \\
\end{eqnarray}
and $N_c=3$ is the color factor.
\begin{figure}
\vspace*{-1cm}
\begin{center}
\includegraphics[width=8cm]{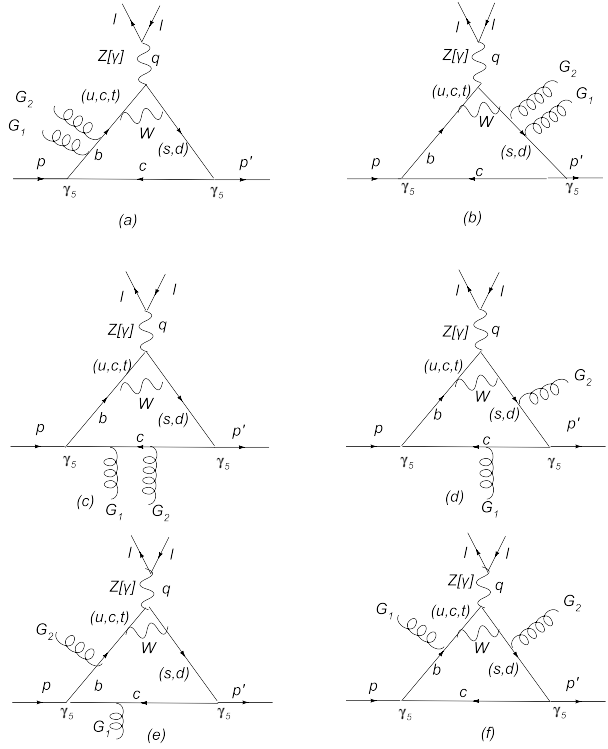}
\end{center}
\caption{Gluon condensate contributions to $B_c \rar
(D,D_s)l^{+}l^{-}/\nu\bar{\nu}$ transitions  } \label{fig2}
\end{figure}

Now as first correction to the non-perturbative part of the
corrolator,  we calculate the gluon condensate contributions (see
diagrams in Fig. 2). The calculations proceed the  same as
\cite{azizi3} ( see also \cite{Aliev3,azizi1,azizi2,R7323}) and  the
Fock--Schwinger fixed--point gauge \cite{R7320,R7321,R7322}, $x^\mu
G_\mu^a = 0$, where $G_\mu^a$ is the gluon field is used. In
calculations, the following type of integrals are encountered:
\begin{eqnarray}  \label{e7323}
I_0[a,b,c] \!\!\! &=& \!\!\! \int \frac{d^4k}{(2 \pi)^4} \frac{1}{\left[
k^2-m_b^2 \right]^a \left[ (p+k)^2-m_c^2 \right]^b \left[ (p^%
\prime+k)^2-m_{q_{_i}}^2\right]^c}~,  \nonumber \\
\nonumber \\
I_\mu[a,b,c] \!\!\! &=& \!\!\! \int \frac{d^4k}{(2 \pi)^4} \frac{k_\mu}{%
\left[ k^2-m_b^2 \right]^a \left[ (p+k)^2-m_c^2 \right]^b \left[
(p^\prime+k)^2-m_{q_{_i}}^2\right]^c}~.  \nonumber \\
\end{eqnarray}

Performing integration over loop momentum  and applying double Borel
transformations with respect to the  $p^2$ and $p^{\prime 2}$, we
obtain  the Borel transformed form of the integrals as follows:
\begin{eqnarray}  \label{e7326}
\hat{I}_0(a,b,c) \!\!\! &=& \!\!\! \frac{(-1)^{a+b+c}}{16 \pi^2\,\Gamma(a)
\Gamma(b) \Gamma(c)} (M_1^2)^{2-a-b} (M_2^2)^{2-a-c} \, \mathcal{U}%
_0(a+b+c-4,1-c-b)~,  \nonumber \\
\nonumber \\
\hat{I}_\mu(a,b,c) \!\!\! &=& \!\!\! \frac{1}{2} \Big[\hat{I}_1(a,b,c) +
\hat{I}_2(a,b,c)\Big] \mathcal{P}_\mu + \frac{1}{2} \Big[\hat{I}_1(a,b,c) -
\hat{I}_2(a,b,c)\Big] q_\mu~,
\end{eqnarray}
where
\begin{eqnarray}  \label{e7327}
\hat{I}_1(a,b,c) \!\!\! &=& \!\!\! i \frac{(-1)^{a+b+c+1}}{16
\pi^2\,\Gamma(a) \Gamma(b) \Gamma(c)} (M_1^2)^{2-a-b} (M_2^2)^{3-a-c} \,
\mathcal{U}_0(a+b+c-5,1-c-b)~,  \nonumber \\
\nonumber \\
\hat{I}_2(a,b,c) \!\!\! &=& \!\!\! i \frac{(-1)^{a+b+c+1}}{16
\pi^2\,\Gamma(a) \Gamma(b) \Gamma(c)} (M_1^2)^{3-a-b} (M_2^2)^{2-a-c} \,
\mathcal{U}_0(a+b+c-5,1-c-b)~,  \nonumber \\
\nonumber \\
\end{eqnarray}
Hat in Eq. (\ref{e7326})  denotes  the double Borel transformed form
of integrals. $M_1^2$ and $M_2^2$ are the Borel parameters in the
$s$ and $s^\prime$ channels, respectively, and the function ${\cal
U}_0(a,b)$ is defined as:
\begin{eqnarray}
\mathcal{U}_0(a,b) = \int_0^\infty dy (y+M_1^2+M_2^2)^a y^b \,exp\left[ -%
\frac{B_{-1}}{y} - B_0 - B_1 y \right]~,  \nonumber
\end{eqnarray}
where
\begin{eqnarray}  \label{e7328}
B_{-1} \!\!\! &=& \!\!\! \frac{1}{M_1^2M_2^2}
\left[m_{q_{_i}}^2M_1^4+m_b^2
M_2^4 + M_2^2M_1^2 (m_b^2+m_{q_{_i}}^2 + Q^2) \right] ~,  \nonumber \\
B_0 \!\!\! &=& \!\!\! \frac{1}{M_1^2 M_2^2} \left[
(m_{q_{_i}}^2+m_c^2) M_1^2
+ M_2^2 (m_b^2+m_c^2) \right] ~,  \nonumber \\
B_{1} \!\!\! &=& \!\!\! \frac{m_c^2}{M_1^2 M_2^2}~.
\end{eqnarray}
After straightforward but lengthy  calculations, we get the
following results for the gluon condensate contributions:
\begin{equation}
\Pi_{i}^{\la G^2 \ra}=i \lla \frac{\alpha_s}{\pi} G^2 \rra
\frac{C_{i}}{6},
\end{equation}
where the explicit expressions for $C_{i}$   are given in
appendix--A.

 Next step is to   apply the  Borel
transformations with respect to the $p^2$ ($p^2\rightarrow
M_{1}^{2}$) and $p'^2$ ($p^{\prime 2}\rightarrow M_{2}^{2}$) on the
phenomenological as well as the perturbative  parts of the
correlation  function, continuum subtraction   and equate these two
representations of the correlator. The following sum rules for the
form factors
 $f_{+}~,f_{-}$ and $f_{T}$ are derived:
\begin{eqnarray}
f_{+} \!\!\! &=& \!\!\!  \frac{(m_b+m_c) (m_c+m_{q_{i}})}{ f_{B_c}
m_{B_c}^2 f_{P} m _{P}^{2}}
e^{m_{B_c}^2/M_1^2} e^{m_{P}^2/M_2^2}  \nonumber \\
&\times& \!\!\! \frac{1}{4 \pi^2} \Bigg\{
\int_{(m_c+m_{q})^2}^{s_0^\prime} ds^\prime
\int_{f_{-}(s')}^{min(s_0,f_{+}(s'))}ds \rho_{+}^{V}
(s,s^\prime,Q^2) e^{-s/M_1^2}
e^{-s^\prime/M_2^2} - iM_1^2M_2^2 \left< \frac{\alpha_s}{\pi} G^2 \right> \frac{C_{+}}{%
6} \Bigg\}~,\nonumber \\
\end{eqnarray}
\begin{eqnarray}
f_{-} \!\!\! &=& \!\!\!  \frac{(m_b+m_c) (m_c+m_{q_{i}})}{ f_{B_c}
m_{B_c}^2 f_{P} m _{P}^{2}}
e^{m_{B_c}^2/M_1^2} e^{m_{P}^2/M_2^2}  \nonumber \\
&\times& \!\!\! \frac{1}{4 \pi^2} \Bigg\{
\int_{(m_c+m_{q})^2}^{s_0^\prime} ds^\prime
\int_{f_{-}(s')}^{min(s_0,f_{+}(s'))}ds \rho_{-}^{V}
(s,s^\prime,Q^2) e^{-s/M_1^2}
e^{-s^\prime/M_2^2} - iM_1^2M_2^2 \left< \frac{\alpha_s}{\pi} G^2 \right> \frac{C_{-}}{%
6} \Bigg\}~,\nonumber \\
\end{eqnarray}
\begin{eqnarray}
f_{T} \!\!\! &=& \!\!\! \frac{(m_b+m_c)(m_c+m_{q_{i}}) }{ f_{B_c}
m_{B_c}^2 f_{P} m _{P}^{2}(m_{B_{c}}-m_{P})} e^{m_{B_c}^2/M_1^2}
e^{m_{P}^2/M_2^2}\nonumber \\
&\times&  \!\!\! \frac{1}{4 \pi^2} \Bigg\{
\int_{(m_c+m_{q})^2}^{s_0^\prime} ds^\prime
\int_{f_{-}(s')}^{min(s_0,f_{+}(s'))}ds \rho_{T}^{T}
(s,s^\prime,Q^2) e^{-s/M_1^2}
e^{-s^\prime/M_2^2} - iM_1^2M_2^2 \left< \frac{\alpha_s}{\pi} G^2 \right> \frac{C_{T}}{%
6} \Bigg\}~.\nonumber \\
\end{eqnarray}
where $s_0$ and $s_0^\prime$ are the continuum thresholds   and
$s=f_{\pm}(s')$ in the lower and upper limit of the integral  over
$s$ are obtained from inequality ( \ref{e7318}). The
$min(s_0,f_{+}(s'))$ means that for  each value of the $q^{2}$, the
smaller one between $s_{0}$ and $f_{+}$  is selected. In above
equations, in order to subtract the contributions of the higher
states and the continuum the quark-hadron duality assumption is also
 used
\begin{eqnarray}
\rho^{higher states}(s,s') = \rho^{OPE}(s,s') \theta(s-s_0)
\theta(s'-s'_0).
\end{eqnarray}

 At the end of this
section, we would like to present the differential decay width of
$B_c \rightarrow Pl^{+}l^{-}/\nu\bar{\nu}$ decays. Using the
 parametrization of these transitions in terms of form
factors and amplitude in Eq.(\ref{e2}), we get

\begin{equation}  \label{2abu}
\frac{d{\Gamma}}{dQ^2}(B_{c}^{\pm} \rightarrow P^\pm {\nu} \bar \nu) =\frac{%
G_{_F}^2{\alpha}^2}{2^8{\pi}^5}{\mid}V_{tq_{_i}}V^*_{tb}{\mid}^2 \phi_{P}%
^{3/2}(1,r_{_P},s)m_{B_{c}}^3{\mid}C_{10}{\mid}^2{\mid}f_+ (Q^2){\mid}%
^2 ,  \label{(8)}
\end{equation}

where, $\phi_{P}(1,r_{_P},s)$ is the usual triangle function
\[
{\phi_{_P}}(1,r_{_P},s)=1+r_{_P}^2 +s^2-2r_{_P}-2s-2r_{_P}s ,\quad
~~
\]
\[
~~\mathrm{with}~~ r_{_P}=\frac{m_{_P}^2}{m_{B_{c}}^2},\quad s=-\frac{Q^2}{%
m_{B_{c}}^2} ,
\]
and
\begin{eqnarray}\label{check_1}
    \frac{d\Gamma  }{dQ^2}\left( B_c^{\pm}\rightarrow P^\pm l^+l^-\right)
        &=&\frac{G_{_F}^2{\mid}V_{tq_{_i}}V^*_{tb}{\mid}^2 m_{B_c}^3\alpha^2}
        {3\cdot 2^9\pi ^5}v \phi_{_P}^{1\over 2}(1,r_{P},s)\left[ \left(1+\frac{2t}
        {s}\right) \phi_{_P}(1,r_{P},s)\alpha _1+12t \beta
        _1\right]\,,\nonumber\\
\end{eqnarray}
where   $t= m_{l}^2/ m_{B_c}^2 $ and  the expressions of
 $\alpha_1$ and $\beta_1$ and $v$
are given as:
\begin{eqnarray*}
v&=&\sqrt{1+\frac{4m_{l}^2}{Q^2}},
\\
 \alpha_1 &=& \biggl|C_{9}^{\rm eff}\,f_{+}(Q^2) +\frac{2\,{\hat
m}_b\,C_{7}^{\rm eff}\,f_{T}(Q^2)}{1+\sqrt{r_{P}}}\biggr|^{2}
       +|C_{10}f_{+}(Q^2)|^{2} ~,
  \\
    \beta_1 &=& |C_{10}|^{2}\biggl[ \biggl(
       1+r_{P}-{s\over 2}\biggr) |f_{+}(Q^2)|^{2}+\biggl( 1-r_{P}\biggr)
       {\rm Re}(f_{+}(Q^2)f_{-}^{*}(Q^2))+\frac{1}{2}s|f_{-}(Q^2)|^{2}\biggr]\,,
\\
\end{eqnarray*}
where ${\hat m}_b=m_b/m_{B_{c}}$.
\section{HQET limit of the  form factors  }
In this section, we present the infinite heavy quark mass limit of
the form factors for $B_c \rar (D,D_s)l^{+}l^{-}/\nu\bar{\nu} $
transitions. To this aim, we use the following parametrization (see
also \cite{ming,neubert1,kazem,melahat,kazemazizi}):
 \begin{equation}\label{melau}
 y=\nu\nu'=\frac{m_{B_{c}}^2+m_{D_{q_{i}}}^2-q^2}{2m_{B_{c}}m_{Dq_{i}}}
 \end{equation}
 where $\nu$ and $\nu'$ are
  the four-velocities of the initial and final meson states, respectively  and $y=1$ are
  so called zero recoil limit. Now,  to obtain the y dependent
  expressions of the form factors we define
  $m_{b}\rightarrow\infty$, $m_{c}=\frac{m_{b}}{\sqrt{z}}$, where
  z is given by  $\sqrt{z}=y+\sqrt{y^2-1}$ and we also set the mass of light quarks to zero.
  In this limit the new Borel
  parameters $ T_{1}$ and $ T_{2}$ take the form $T_{1}=M_{1}^{2}/2  m_{b}$ and $T_{2}=M_{2}^{2}/2
  m_{c}$.

  The new continuum thresholds $\nu_{0}$, and
$\nu_{0}'$ are defined as:
\begin{equation}\label{17au}
 \nu_{0}=\frac{s_{0}-m_{b}^2}{m_{b}},~~~~~~
 \nu'_{0}=\frac{s'_{0}-m_{c}^2}{m_{c}},
 \end{equation}
 and the new integration variables become :
 \begin{equation}\label{18au}
 \nu=\frac{s-m_{b}^2}{m_{b}},~~~~~~ \nu'=\frac{s'-m_{c}^2}{m_{c}}.
 \end{equation}
 The leptonic decay constants are rescaled:
 \begin{equation}\label{21au}
\hat{f}_{B_{c}}=\sqrt{m_{b}}
f_{B_{c}},~~~~~~~\hat{f}_{D_{q_{i}}}=\sqrt{m_{c}} f_{D_{q_{i}}}.
\end{equation}
 The correspond expressions for $\hat{I}_0(a,b,c)$, $\hat{I}_1(a,b,c)$ and $\hat{I}_2(a,b,c)$ in this limit are defined as:
\begin{eqnarray}  \label{e7326guy}
\hat{I}_0(a,b,c)^{HQET} \!\!\! &=& \!\!\! \frac{(-1)^{a+b+c}}{16
\pi^2\,\Gamma(a)
\Gamma(b) \Gamma(c)} (T_1)^{2-a-b} (T_2)^{2-a-c} \, \mathcal{U}%
_0^{HQET}(a+b+c-4,1-c-b)~,  \nonumber \\
\hat{I}_1(a,b,c)^{HQET} \!\!\! &=& \!\!\! i \frac{(-1)^{a+b+c+1}}{16
\pi^2\,\Gamma(a) \Gamma(b) \Gamma(c)} (T_1)^{2-a-b} (T_2)^{3-a-c} \,
\mathcal{U}_0^{HQET}(a+b+c-5,1-c-b)~,  \nonumber \\
\nonumber \\
\hat{I}_2(a,b,c)^{HQET} \!\!\! &=& \!\!\! i \frac{(-1)^{a+b+c+1}}{16
\pi^2\,\Gamma(a) \Gamma(b) \Gamma(c)} (T_1)^{3-a-b} (T_2)^{2-a-c} \,
\mathcal{U}_0^{HQET}(a+b+c-5,1-c-b)~,  \nonumber \\
\nonumber \\
\end{eqnarray}
where $T_1$ and $T_2$ are the Borel parameters in the $s$ and
$s^\prime$ channel, respectively, and the function
$\mathcal{U}_0^{HQET}(m,n)$ is defined as
\begin{eqnarray}
\mathcal{U}_0^{HQET}(m,n)=\int _{0}^{\infty }\! ( x+{ T_1}+{ T_2} )
^{m}{x}^{n}\exp\{\hat{A}+\hat{B}+\hat{C}\}{dx},
\end{eqnarray}
with
\begin{eqnarray}
A&=&-{\frac{{{m_{b}}}^{2}{{T_{2}}}^{2}+{T_{1}}\,{T_{2}}\,({{m_{b}}}^{2}+{y}%
)}{{T_{1}}\,{T_{2}}\,x}} \nonumber\\
B&=&-\frac{{T_{2}}\,({{m_{b}}}^{2}+{\frac{{{m_{b}}}^{2}}{z}})+{\frac{{T_{1}}%
\,{{m_{b}}}^{2}}{z}}}{{{T_{1}}}{{T_{2}}}}\nonumber \\
C &=&-{\frac{{{m_{b}}}^{2}x}{{T_{1}}\,{T_{2}}\,z}}.
\end{eqnarray}
 In order to the calculations be easy, the following redefinitions for the form factors are applied
\begin{eqnarray}
\tilde{f}_{i}=f_{i}\{m_{B_{c}}+m_{D_{q_{i}}}\}
\end{eqnarray}
 After some calculations, we obtain the y-dependent
expressions of the form factors as follows:
\begin{eqnarray}\label{heavyp}
\tilde{f}_{+}^{HQET}(y) &=&\frac{1}{32\pi ^{2}\hat{f}_{B_{c}}\hat{f}_{P}}e^{\frac{%
\Lambda }{T_{1}}}e^{\frac{\overline{{\Lambda }}}{T_{2}}}\Bigg\{\frac{3(1+%
\sqrt{z})^{2}}{(-1+\sqrt{z})F(y,z)} \nonumber\\
&&[-1+(1+3y)\sqrt{z}-(2+y+4y^{2})z+(3+2y(2+y))z^{3/2}-4yz^{2}+2z^{5/2}] \nonumber\\
&&\int_{0}^{\nu _{0}}d\nu \int_{0}^{\nu _{0}^{^{\prime }}}d{\nu
}^{^{\prime }}e^{-\frac{\nu }{2T_{1}}}e^{-\frac{{\nu }^{_{^{\prime
}}}}{2T_{2}}}\theta
(2y\nu \nu ^{^{\prime }}-\nu ^{2}-{{\nu }^{_{^{\prime }}}}^{2})\nonumber \\
&&+\lim_{m_{b}\rightarrow \infty }~~\Bigg(i~\frac{5z^2}{24{m_{b}^{5}}}~(1+\sqrt{%
z})\left\langle \frac{\alpha _{s}}{\pi }G^{2}\right\rangle
C_{_{+}}^{HQET}\Bigg)\Bigg\}~,
\end{eqnarray}

\begin{eqnarray}\label{heavym}
\tilde{f}_{-}^{HQET}(y) &=&\frac{1}{32\pi ^{2}\hat{f}_{B_{c}}\hat{f}_{P}}e^{\frac{%
\Lambda }{T_{1}}}e^{\frac{\overline{{\Lambda }}}{T_{2}}}\Bigg\{\frac{-3(1+%
\sqrt{z})^{2}}{(-1+\sqrt{z})F(y,z)}\nonumber \\
&&[-1+(1+3y)\sqrt{z}-(2+5y)z+(7+2y(y-2))z^{3/2}-4(y-1)z^{2}+2z^{5/2}] \nonumber\\
&&\int_{0}^{\nu _{0}}d\nu \int_{0}^{\nu _{0}^{^{\prime }}}d{\nu
}^{^{\prime }}e^{-\frac{\nu }{2T_{1}}}e^{-\frac{{\nu }^{_{^{\prime
}}}}{2T_{2}}}\theta
(2y\nu \nu ^{^{\prime }}-\nu ^{2}-{{\nu }^{_{^{\prime }}}}^{2}) \nonumber\\
&&+\lim_{m_{b}\rightarrow \infty }~~\Bigg(i~\frac{5z^2}{24{m_{b}^{5}}}~(1+\sqrt{%
z})\left\langle \frac{\alpha _{s}}{\pi }G^{2}\right\rangle
C_{_{-}}^{HQET}\Bigg)\Bigg\}~,
\end{eqnarray}

\begin{eqnarray}\label{heavyt}
\tilde{f}_{T}^{HQET}(y) &=&\frac{1}{32\pi ^{2}\hat{f}_{B_{c}}\hat{f}_{P}}e^{\frac{%
\Lambda }{T_{1}}}e^{\frac{\overline{{\Lambda }}}{T_{2}}}\Bigg\{\frac{-3(1+%
\sqrt{z})^{2}}{(-1+\sqrt{z})F(y,z)} \nonumber\\
&&\lbrack
3-(1+9y)\sqrt{z}+(4+y(3+8y))z-((2+y)(1+4y))z^{3/2}+(3+2y(y+2))z^{2}
\nonumber\\
&&-4yz^{5/2}+z^{3}]\int_{0}^{\nu _{0}}d\nu \int_{0}^{\nu _{0}^{^{\prime }}}d{%
\nu }^{^{\prime }}e^{-\frac{\nu }{2T_{1}}}e^{-\frac{{\nu }^{_{^{\prime }}}}{%
2T_{2}}}\theta (2y\nu \nu ^{^{\prime }}-\nu ^{2}-{{\nu }^{_{^{\prime }}}}%
^{2}) \nonumber\\
&&+\lim_{m_{b}\rightarrow \infty }~~\Bigg(i~ \frac{5}{48~z{m_{b}^5}}~(1+\sqrt{z})^2\left\langle \frac{%
\alpha _{s}}{\pi }G^{2}\right\rangle C_{_{T}}^{HQET}\Bigg)\Bigg\}~,
\end{eqnarray}
where
\begin{eqnarray}
F(y,z)=z^{3/4}[1+z+y^2z+z^2-2y\sqrt{z}(1+z)]^{3/2}.
\end{eqnarray}
In the heavy quark limit expressions of the form factors,
$\Lambda=m_{B_{q}}-m_{b}$ and $\bar{\Lambda}=m_{D_{q}^{*}}-m_{c}$.
and the explicit expressions of the coefficients $C_{i}^{HQET}$ are
given in the appendix--B.
\section{Numerical analysis}

This section encompasses our numerical analysis of the form factors $%
f_{+}~,~f_{-}~$~and~$f_{T}$ and their HQET limit,  branching
fractions, comparison of our results with the prediction of the RCQM
and discussion. The sum rules expressions of the form factors depict
that the main input parameters entering  the expressions are gluon
condensate, Wilson coefficients $C_{7}^{eff}$, $C_{9}^{eff}$ and
$C_{10}$ , elements of the CKM matrix $V_{tb}$, $V_{ts}$ and
$V_{td}$, leptonic decay constants; $f_{B_{C}}$, $f_{D}$ and
$f_{D_{s}}$, Borel parameters $M_{1}^2$ and $M_{2}^2$, as well as
the continuum thresholds $s_{0}$ and $s'_{0}$. In further  numerical
analysis, we choose the values of the Gluon condensate, leptonic
decay constants, CKM matrix elements, Wilson coefficients, quark and
meson masses as: $<\frac{\alpha_{s}}{\pi}G^{2}>=0.012~ GeV ^{4}$
\cite{Shifman1},
  $C_7^{eff}=-0.313$, $C_9^{eff}=4.344$, $C_{10}=-4.669$
 \cite{Buras,Bashiry},  $\mid
V_{tb}\mid=0.77^{+0.18}_{-0.24}$, $\mid
V_{ts}\mid=(40.6\pm2.7)\times10^{-3}$ $\mid
V_{td}\mid=(7.4\pm0.8)\times10^{-3}$ \cite {Ceccucci}, $f_{D_{s}}
=274\pm13\pm7
  ~MeV $ \cite{Artuso1},
  $f_{D}=222.6\pm16.7^{+2.8}_{-3.4}~MeV$,
\cite{Artuso2}, $f_{B_{c}} =350\pm25
  ~MeV $ \cite{Colangelo2,Kiselev,Alievy},  $ m_{c}=
 1.25\pm
 0.09~ GeV$, $m_{s}=95\pm25 ~MeV$  ,  $m_{b} =
(4.7\pm
 0.07)~GeV$, $m_{d}=(3-7)~MeV$, $m_{D_{s}}=1.968~GeV$,
 $m_{D}=1.869~GeV$,
   $ m_{B_{C}}=6.258~GeV$ \cite{Yao}, $\Lambda=0.62 GeV$ \cite{huang} and $\overline{\Lambda}=0.86
GeV$\cite{dai}.

The expressions for the form factors  contain also four auxiliary
parameters: Borel mass squares $M_{1}^2$ and $M_{2}^2$ and continuum
threshold $s_{0}$  and $s'_{0}$. These are not physical quantities,
so the physical quantities, form factors, should be independent of
them. The parameters $s_0$ and $s_0^\prime$, which are the continuum
thresholds of $B_c$ and $P$ mesons, respectively, are
 determined from the conditions that
guarantees the sum rules to have the best stability in the allowed
$M_1^2$ and $M_2^2$ region. The values of continuum thresholds
calculated from the two--point QCD sum rules are taken to be
$s_0=(45-50)~GeV^2$ and $s_0^\prime=(6-8)~GeV^2$
\cite{Aliev1,Shifman1,Colangelo1}.  The working regions for $M_1^2$
and $M_2^2$ are determined by requiring that not only contributions
of the   higher states and continuum are effectively suppressed, but
the gluon condensate contributions are small, which guarantees that
the contributions of higher dimensional operators are small. Both
conditions are satisfied in the  regions $10~GeV^2 \le M_1^2 \le
25~GeV^2$ and $4~GeV^2 \le M_2^2 \le 10~GeV^2$.

The dependence of
the form factors  $%
f_{+}~,~f_{-}~$~and~$f_{T}$ on $M_1^2$ and $M_2^2$ for $B_c
\rightarrow D_s l^{+}l^{-}/\nu\bar{\nu}$ are shown in Figs. 3, 4 and
5, respectively. The figures 6, 7, and 8 also depict the dependence
of the  the form factors on Borel mass parameters for for $B_c
\rightarrow D l^{+}l^{-}/\nu\bar{\nu}$. This figures show a good
stability of the form factors with respect to the Borel mass
parameters in the working regions. Our numerical analysis  shows
that   the contribution of the non-perturbative part (the gluon
condensate diagrams ) is about $8^0/_{0}$ of the total and the main
contribution comes from the perturbative part of the form factors.

The values of  the form factors at $q^2=0$ are  shown in Table 1:
\begin{table}[h]
\centering
\begin{tabular}{|c||c|c|}
\hline
& $B_c \rightarrow D $ & $B_c \rightarrow D_s $ \\
\cline{1-3} \hline\hline$f_{+}(l^{+}l^{-}/\nu\bar{\nu})$ &$ 0.22 \pm0.045$& $0.16 \pm0.032$\\
\cline{1-3} $f_{-}(l^{+}l^{-}/\nu\bar{\nu})$&$-0.29\pm0.056$&
$-0.18\pm0.038 $\\ \cline{1-3} $f_{T}(l^{+}l^{-})$ &$-0.27\pm0.054$
&$-0.19\pm0.040$\\\cline{1-3}
\end{tabular}
\vspace{0.8cm} \caption{The values of  the form factors at $q^2=0$}
\label{tab:1}
\end{table}
 The sum rules for the form factors are truncated at about $2 ~GeV^2$ below the perturbative cut, so  to extend our results to the full physical region, we
look for parametrization of the form factors in such a way that in
the region $0 \leq q^2 \leq 19.26~(18.41)~ GeV^2$ for $D(D_{s})$,
this parametrization coincides with the sum rules prediction. Our
numerical calculations shows that the sufficient parametrization
 of the form factors with respect to $q^2$ is as follows:%
\newline
\begin{equation}  \label{17au}
f_{i}(q^2)=\frac{f_{i}(0)}{1+ \alpha\hat{q}+ \beta\hat{q}%
^2}
\end{equation}
where $\hat{q}=q^2/m_{B_{c}}^2$. The values of the parameters $%
f_{i}(0), ~\alpha$ and  $\beta$ are given in the Tables 2, 3.
\begin{table}[h]
\centering
\begin{tabular}{|c||c|c|c|}
\hline & f(0) & $ \alpha$ & $ \beta$
\\ \cline{1-4} \hline\hline$f_{+}(l^{+}l^{-}/\nu\bar{\nu})$ &  0.22 & -1.10 & -2.48   \\
\cline{1-4} $f_{-}(l^{+}l^{-}/\nu\bar{\nu})$ &-0.29 & -0.63& -4.06
 \\ \cline{1-4} $f_{T}(l^{+}l^{-})$ & -0.27 & -0.72 & -3.24  \\
\cline{1-4}
\end{tabular}
\vspace{0.8cm} \caption{Parameters appearing in the form factors of
the $B_{c}\rightarrow D l^{+}l^{-}/\nu\bar{\nu}$ decay  for
$M_{1}^2=15~GeV^2$, $ M_{2}^2=8~GeV^2$.} \label{tab:1}
\end{table}
\begin{table}[h]
\centering
\begin{tabular}{|c||c|c|c|}
\hline & f(0) & $ \alpha$ & $ \beta$
\\ \cline{1-4} \hline \hline $f_{+}(l^{+}l^{-}/\nu\bar{\nu})$ &  0.16 & -1.55 & -2.80 \\ \cline{1-4} $f_{-}(l^{+}l^{-}/\nu\bar{\nu})$
&-0.18 & -0.77& -6.71 \\ \cline{1-4} $f_{T}(l^{+}l^{-})$ & -0.19 & -1.43 &-3.06  \\
\cline{1-4}
\end{tabular}
\vspace{0.8cm} \caption{Parameters appearing in the form factors of
the $B_{c}\rightarrow D_sl^{+}l^{-}/\nu\bar{\nu}$ decay  for
$M_{1}^2=15~GeV^2$, $ M_{2}^2=8~GeV^2$.} \label{tab:1}
\end{table}
The errors are estimated by the variation of the Borel parameters $M_1^2$
and $M_2^2$, the variation of the continuum thresholds $s_0$ and $s_0^\prime$%
, the variation of $b$ and $c$ quark masses and leptonic decay constants $%
f_{B_c}$ and $f_{D,(D_s)}$. The main uncertainty comes from the
thresholds and the decay constants, which is about $\sim 18\%$ of
the central value, while the other uncertainties are small,
constituting a few percent.

Now, we compare the extrapolation values for the form factors and
their HQET  values obtained from Eqs. (\ref{heavyp}-\ref{heavyt}) in
Tables 4 and 5 for $B_{c}\rightarrow D l^{+}l^{-}/\nu\bar{\nu}$ and
$B_{c}\rightarrow D_{s} l^{+}l^{-}/\nu\bar{\nu}$, respectively.
\begin{table}[h]
\centering
\begin{tabular}{|c|c|c|c|c|c|c|c|c|c|c|}
\hline $y$ &1& 1.1& 1.2& 1.3& 1.4& 1.5& 1.6& 1.7& 1.8  \\
\cline{1-10}$q^2$ &19.26& 16.93&14.59 &12.25&9.91&7.57&5.23&2.89&0.55 \\
\cline{1-10}$f_{+}(q^2)$ &2.19 & 1.36&0.88 &0.56 &0.36&0.29&0.27&0.24&0.23 \\
\cline{1-10}$f_{-}(q^2)$ &-3.01& -1.93&-1.20 &-0.75 &-0.52&-0.39&-0.33&-0.32&-0.31 \\
\cline{1-10}$f_{_{T}}(q^2)$ &-2.52&-1.53 & -1.12&-0.70 &-0.49 &-0.37&-0.31&-0.29&-0.28\\
\cline{1-10}$f^{HQET}_{+}(y)$& ?&1.35 & 0.50&0.29&0.20 &0.15&0.12&0.10&0.08 \\
\cline{1-10}$f^{HQET}_{-}(y)$ &?&-1.90 & -0.75&-0.44 &-0.30 &-0.22&-0.18&-0.15&-0.12 \\
\cline{1-10}$f^{{HQET}}_{_{T}}(y)$ &?&-1.51 & -0.58&-0.33 &-0.23&-0.17&-0.14&-0.11&-0.10 \\
\cline{1-10}
\end{tabular}
\vspace{0.10cm} \caption{The comparison of the extrapolation values
for the form factors and their HQET limit for  $B_c \rightarrow D
l^{+}l^{-}$ at $M_1^{2}=15~GeV^2,~M_2^{2}=8~GeV^2$ and corresponding
$T_1=1.6~GeV,T_2=3.2~GeV$.} \label{tab:1}
\end{table}

\begin{table}[h]
\centering
\begin{tabular}{|c|c|c|c|c|c|c|c|c|c|c|}
\hline $y$ &1& 1.1& 1.2& 1.3& 1.4& 1.5& 1.6& 1.7 \\
\cline{1-9}$q^2$ &18.41 & 15.94&13.48 &11.02 &8.55&6.09&3.63&1.16\\
\cline{1-9}$f_{+}(q^2)$ &2.17 & 1.12&0.79 &0.53 &0.31&0.22&0.18&0.17\\
\cline{1-9}$f_{-}(q^2)$ &-2.50 & -1.53&-0.79 &-0.43 &-0.29&-0.24&-0.23&-0.22\\
\cline{1-9}$f_{_{T}}(q^2)$ &-2.25 &- 1.23&-0.70&-0.37 &-0.27 &-0.23&-0.21&-0.20\\
\cline{1-9}$f^{HQET}_{+}(y)$& ?&1.08 & 0.41&0.24&0.16 &0.12&0.10&0.08 \\
\cline{1-9}$f^{HQET}_{-}(y)$ &?&-1.52 & -0.60&-0.35 &-0.24 &-0.18&-0.14&-0.12\\
\cline{1-9}$f^{HQET}_{_{T}}(y)$ &?&-1.22 & -0.46&-0.27 &-0.18&-0.14&-0.11&-0.10 \\
\cline{1-9}
\end{tabular}
\vspace{0.10cm} \caption{The comparison of the extrapolation values
for the form factors and their HQET limit for  $B_c \rightarrow
D_{s} l^{+}l^{-}$ at $M_1^{2}=15~GeV^2,~M_2^{2}=8~GeV^2$ and
corresponding $T_1=1.6~GeV,T_2=3.2~GeV.$} \label{tab:1}
\end{table}
At the $y=1$ called the zero recoil limit, the HQET limit of the
form factors are not finite and at this value, we can determine only
the ratio of the form factors. For other values of y and
corresponding $q^2$, the behavior of the form factors and their HQET
values are the same, i.e., when y increases ($q^2$ decreases) both
the form factors and their HQET values decrease. Moreover, at high
$q^2$ values, the form factors and their HQET values are close to
each other while at low $q^2$, the form factor values are about 2-3
times greater than that of their HQET limit.

At the end of this section we would like to present the values of
the branching ratios. Integrating Eqs. (\ref{2abu}) and
(\ref{check_1}) over $q^2$ in the whole physical region and using
the total mean life time $\tau \simeq 0.46~ps$ of $B_c$ meson \cite
{R7326}, the branching ratio of the $B_c \rightarrow P(D,D_s)
l^{+}l^{-}/\nu\bar{\nu}$ decays are obtained as Table 6. This Table
also includes a comparison of our results with the prediction of the
RCQM. This Table presents a good agreement between two models
especially when the errors are taken into account. Any
experimentally measurements on the branching fractions of these
decays and those comparisons with the results of the
phenomenological models like QCD sum rules could give valuable
information about the nature of the $D_{s}$ meson and strong
interactions inside it.

In summary, we investigated the $B_c \rightarrow P(D,D_s)
l^{+}l^{-}/\nu\bar{\nu}$ channels and computed the relevant form
factors and their HQET limits considering the gluon condensate
corrections. We also evaluated the total decay width and the
branching fractions of those decays and compared our results with
the predictions of the RCQM. Detection of these channels and their
comparison with the phenomenological models like QCD sum rules could
give useful information about the structure of the $D_{s}$ meson.
\begin{table}[h]
\centering
\begin{tabular}{|c||c|c|}
\hline decay & our results & RCQM\cite{tt1} \\ \cline{1-3}
\hline\hline$B_c
\rightarrow D \nu\bar{\nu}$ & $(3.48 \pm 0.71)\times10^{-8}$ & $3.28\times10^{-8}$ \\
\cline{1-3}
$B_c \rightarrow D_s \nu\bar{\nu}$ & $(0.49\pm 0.12)\times10^{-6}$& $0.7\times10^{-6}$ \\
\cline{1-3} $B_c \rightarrow D e^+ e^-$ &$(1.34\pm 0.25)\times10^{-8}$& - \\
\cline{1-3} $B_c \rightarrow D_s e^+ e^-$ & $(1.47\pm 0.32)\times10^{-7}$& -\\
\cline{1-3} $B_c \rightarrow D \mu^+\mu^-$ &
$(0.31\pm0.06)\times10^{-8}$ & $0.44\times10^{-8}$
\\ \cline{1-3} $B_c
\rightarrow D_s \mu^+\mu^-$ &$ (0.61\pm 0.15)\times10^{-7}$ & $0.97 \times10^{-7}$\\
\cline{1-3} $B_c \rightarrow D \tau^+\tau^- $ &$(0.13\pm 0.03)\times10^{-8}$&$ 0.11\times10^{-8}$\\
\cline{1-3} $B_c \rightarrow D_s \tau^+\tau^-$ & $(0.23\pm
0.05)\times10^{-7}$ &$ 0.22\times10^{-7}$\\ \cline{1-3}
\end{tabular}
\vspace{0.8cm} \caption{Values for the branching fractions  of the
$B_{c}\rightarrow P(D,D_s)l^{+}l^{-}/\nu\bar{\nu}$ decays and their
comparison  with the predictions of the RCQM \cite{tt1}}
\label{tab:1}
\end{table}

\newpage

\section*{Acknowledgments}
The authors would like to thank T. M. Aliev and A. Ozpineci for
their useful discussions. One of the authors (K. Azizi) thanks
Turkish Scientific and Research Council (TUBITAK) for their partial
financial support.

\appendix

\appendix

\begin{center}
{\Large \textbf{Appendix--A}}
\end{center}


\setcounter{equation}{0} \renewcommand{\theequation}{C.\arabic{equation}}
\setcounter{section}{0} \setcounter{table}{0}

In this appendix,  the explicit expressions of the coefficients of
the gluon condensate  entering  the sum rules of the form factors
$f_{+} ,f_{-}$ and $f_{T}$ are given.
\newline
\newline
\newline
\begin{eqnarray*}
&&C_{+} =-5\,I_{{1}}(3,2,2){\mathit{m_{c}}}^{6}-5\,I_{{2}}(3,2,2){\mathit{%
m_{c}}}^{6}-5\,I_{{0}}(3,2,2){\mathit{m_{c}}}^{6} \\
&&+5\,I_{{2}}(3,2,2){\mathit{m_{c}}}^{5}\mathit{m_{b}}+5\,I_{{1}}(3,2,2){%
\mathit{m_{c}}}^{5}\mathit{m_{b}}+5\,I_{{0}}(3,2,2){\mathit{m_{c}}}^{5}%
\mathit{m_{b}} \\
&&+5\,I_{{2}}(3,2,2){\mathit{m_{c}}}^{4}{\mathit{m_{b}}}^{2}+5\,I_{{1}%
}(3,2,2){\mathit{m_{c}}}^{4}{\mathit{m_{b}}}^{2}-5\,I_{{1}}(3,2,2){\mathit{%
m_{c}}}^{3}{\mathit{m_{b}}}^{3} \\
&&-5\,I_{{0}}(3,2,2){\mathit{m_{c}}}^{3}{\mathit{m_{b}}}^{3}-5\,I_{{2}%
}(3,2,2){\mathit{m_{c}}}^{3}{\mathit{m_{b}}}^{3}-5\,I_{{2}}(3,2,1){\mathit{%
m_{c}}}^{4} \\
&&+15\,I_{1}^{[0,1]}(3,2,2){\mathit{m_{c}}}^{4}-15\,I_{{1}}(2,2,2){\mathit{%
m_{c}}}^{4}+15\,I_{2}^{[0,1]}(3,2,2){\mathit{m_{c}}}^{4} \\
&&-5\,I_{{2}}(3,1,2){\mathit{m_{c}}}^{4}-5\,I_{{1}}(3,1,2){\mathit{m_{c}}}%
^{4}-15\,I_{{2}}(4,1,1){\mathit{m_{c}}}^{4} \\
&&-5\,I_{{1}}(3,2,1){\mathit{m_{c}}}^{4}-15\,I_{{0}}(4,1,1){\mathit{m_{c}}}%
^{4}-15\,I_{{0}}(2,2,2){\mathit{m_{c}}}^{4} \\
&&+15\,I_{0}^{[0,1]}(3,2,2){\mathit{m_{c}}}^{4}-15\,I_{{1}}(4,1,1){\mathit{%
m_{c}}}^{4}-15\,I_{{2}}(2,2,2){\mathit{m_{c}}}^{4} \\
&&+5\,I_{{1}}(3,1,2){\mathit{m_{c}}}^{3}\mathit{m_{b}}-10%
\,I_{0}^{[0,1]}(3,2,2){\mathit{m_{c}}}^{3}\mathit{m_{b}}+10\,I_{{1}}(2,2,2){%
\mathit{m_{c}}}^{3}\mathit{m_{b}} \\
&&+10\,I_{{0}}(3,2,1){\mathit{m_{c}}}^{3}\mathit{m_{b}}-10%
\,I_{2}^{[0,1]}(3,2,2){\mathit{m_{c}}}^{3}\mathit{m_{b}}+5\,I_{{0}}(3,1,2){%
\mathit{m_{c}}}^{3}\mathit{m_{b}} \\
&&-10\,I_{{2}}(2,3,1){\mathit{m_{c}}}^{3}\mathit{m_{b}}+15\,I_{{0}}(4,1,1){%
\mathit{m_{c}}}^{3}\mathit{m_{b}}+5\,I_{{1}}(3,2,1){\mathit{m_{c}}}^{3}%
\mathit{m_{b}} \\
&&+15\,I_{{1}}(4,1,1){\mathit{m_{c}}}^{3}\mathit{m_{b}}-10\,I_{{1}}(2,3,1){%
\mathit{m_{c}}}^{3}\mathit{m_{b}}+15\,I_{{2}}(4,1,1){\mathit{m_{c}}}^{3}%
\mathit{m_{b}} \\
&&+10\,I_{{2}}(2,2,2){\mathit{m_{c}}}^{3}\mathit{m_{b}}-10\,I_{{0}}(2,3,1){%
\mathit{m_{c}}}^{3}\mathit{m_{b}}-10\,I_{1}^{[0,1]}(3,2,2){\mathit{m_{c}}}%
^{3}\mathit{m_{b}} \\
&&+5\,I_{{2}}(3,1,2){\mathit{m_{c}}}^{3}\mathit{m_{b}}+10\,I_{{0}}(2,2,2){%
\mathit{m_{c}}}^{3}\mathit{m_{b}}+5\,I_{{2}}(3,2,1){\mathit{m_{c}}}^{3}%
\mathit{m_{b}} \\
&&+5\,I_{0}^{[0,1]}(3,2,2){\mathit{m_{c}}}^{2}{\mathit{m_{b}}}^{2}+30\,I_{{0}%
}(1,4,1){\mathit{m_{c}}}^{2}{\mathit{m_{b}}}^{2}-5\,I_{{0}}(2,2,2){\mathit{%
m_{c}}}^{2}{\mathit{m_{b}}}^{2} \\
&&-10\,I_{{1}}(3,2,1){\mathit{m_{c}}}^{2}{\mathit{m_{b}}}^{2}+30\,I_{{2}%
}(1,4,1){\mathit{m_{c}}}^{2}{\mathit{m_{b}}}^{2}-10\,I_{{2}}(3,2,1){\mathit{%
m_{c}}}^{2}{\mathit{m_{b}}}^{2} \\
&&+30\,I_{{1}}(1,4,1){\mathit{m_{c}}}^{2}{\mathit{m_{b}}}^{2}-5%
\,I_{0}^{[0,1]}(3,2,2)\mathit{m_{c}}\,{\mathit{m_{b}}}^{3}-30\,I_{{0}}(1,4,1)%
\mathit{m_{c}}\,{\mathit{m_{b}}}^{3} \\
&&+10\,I_{{2}}(2,3,1)\mathit{m_{c}}\,{\mathit{m_{b}}}^{3}+10\,I_{{1}}(2,3,1)%
\mathit{m_{c}}\,{\mathit{m_{b}}}^{3}-5\,I_{1}^{[0,1]}(3,2,2)\mathit{m_{c}}\,{%
\mathit{m_{b}}}^{3} \\
&&-5\,I_{2}^{[0,1]}(3,2,2)\mathit{m_{c}}\,{\mathit{m_{b}}}^{3}+10\,I_{{2}%
}(3,2,1)\mathit{m_{c}}\,{\mathit{m_{b}}}^{3}-30\,I_{{1}}(1,4,1)\mathit{m_{c}}%
\,{\mathit{m_{b}}}^{3} \\
&&+10\,I_{{1}}(3,2,1)\mathit{m_{c}}\,{\mathit{m_{b}}}^{3}-30\,I_{{2}}(1,4,1)%
\mathit{m_{c}}\,{\mathit{m_{b}}}^{3}+15\,I_{{0}}(1,4,1){\mathit{m_{b}}}^{4}
\\
&&-5\,I_{{0}}(3,2,1){\mathit{m_{b}}}^{4}+15\,I_{1}^{[0,1]}(3,2,1){\mathit{%
m_{c}}}^{2}-5\,I_{{0}}(3,1,1){\mathit{m_{c}}}^{2} \\
&&-5\,I_{{0}}(2,2,1){\mathit{m_{c}}}^{2}-5\,I_{{0}}(2,1,2){\mathit{m_{c}}}%
^{2}+30\,I_{1}^{[0,1]}(2,2,2){\mathit{m_{c}}}^{2} \\
&&+15\,I_{0}^{[0,1]}(4,1,1){\mathit{m_{c}}}^{2}+15\,I_{2}^{[0,1]}(3,2,1){%
\mathit{m_{c}}}^{2}+20\,I_{0}^{[0,1]}(3,2,1){\mathit{m_{c}}}^{2} \\
&&-15\,I_{2}^{[0,2]})(3,2,2){\mathit{m_{c}}}^{2}+20\,I_{0}^{[0,1]}(3,1,2){%
\mathit{m_{c}}}^{2}+15\,I_{2}^{[0,1]}(4,1,1){\mathit{m_{c}}}^{2} \\
&&+15\,I_{1}^{[0,1]}(4,1,1){\mathit{m_{c}}}^{2}-15\,I_{1}^{[0,2]}(3,2,2){%
\mathit{m_{c}}}^{2}+15\,I_{2}^{[0,1]}(3,1,2){\mathit{m_{c}}}^{2} \\
&&-10\,I_{{1}}(1,2,2){\mathit{m_{c}}}^{2}-10\,I_{{0}}(1,2,2){\mathit{m_{c}}}%
^{2}+30\,I_{0}^{[0,1]}(2,2,2){\mathit{m_{c}}}^{2} \\
&&+30\,(I_{2}^{[0,1]}(2,2,2){\mathit{m_{c}}}^{2}-10\,I_{{2}}(1,2,2){\mathit{%
m_{c}}}^{2}+15\,I_{1}^{[0,1]}(3,1,2){\mathit{m_{c}}}^{2} \\
&&-15\,I_{0}^{[0,2]}(3,2,2){\mathit{m_{c}}}^{2}-25\,I_{{0}}(2,2,1)\mathit{%
m_{c}}\,\mathit{m_{b}}-5\,I_{{2}}(2,1,2)\mathit{m_{c}}\,\mathit{m_{b}} \\
&&-10\,I_{0}^{[0,1]}(3,2,1)\mathit{m_{c}}\,\mathit{m_{b}}-20\,I_{{1}}(2,2,1)%
\mathit{m_{c}}\,\mathit{m_{b}}-40\,I_{{0}}(1,3,1)\mathit{m_{c}}\,\mathit{%
m_{b}} \\
&&-20\,I_{{2}}(2,2,1)\mathit{m_{c}}\,\mathit{m_{b}}-10\,I_{1}^{[0,1]}(2,2,2)%
\mathit{m_{c}}\,\mathit{m_{b}}-10\,I_{{1}}(1,2,2)\mathit{m_{c}}\,\mathit{%
m_{b}} \\
&&-10\,I_{2}^{[0,1]}(2,2,2)\mathit{m_{c}}\,\mathit{m_{b}}-40\,I_{{1}}(1,3,1)%
\mathit{m_{c}}\,\mathit{m_{b}}-5\,I_{{1}}(2,1,2)\mathit{m_{c}}\,\mathit{m_{b}%
} \\
&&-5\,I_{2}^{[0,1]}(3,2,1)\mathit{m_{c}}\,\mathit{m_{b}}-10%
\,I_{0}^{[0,1]}(2,2,2)\mathit{m_{c}}\,\mathit{m_{b}}-15\,I_{1}^{[0,1]}(3,1,2)%
\mathit{m_{c}}\,\mathit{m_{b}} \\
&&-5\,I_{{0}}(2,1,2)\mathit{m_{c}}\,\mathit{m_{b}}+5\,I_{1}^{[0,2]}(3,2,2)%
\mathit{m_{c}}\,\mathit{m_{b}}-15\,I_{0}^{[0,1]}(3,1,2)\mathit{m_{c}}\,%
\mathit{m_{b}} \\
&&+10\,I_{0}^{[0,1]}(2,3,1)\mathit{m_{c}}\,\mathit{m_{b}}-10\,I_{{2}}(1,2,2)%
\mathit{m_{c}}\,\mathit{m_{b}}-10\,I_{{0}}(1,2,2)\mathit{m_{c}}\,\mathit{%
m_{b}} \\
&&-15\,I_{2}^{[0,1]}(3,1,2)\mathit{m_{c}}\,\mathit{m_{b}}+5%
\,I_{2}^{[0,2]}(3,2,2)\mathit{m_{c}}\,\mathit{m_{b}}-5\,I_{1}^{[0,1]}(3,2,1)%
\mathit{m_{c}}\,\mathit{m_{b}} \\
&&+5\,I_{0}^{[0,2]}(3,2,2)\mathit{m_{c}}\,\mathit{m_{b}}-40\,I_{{2}}(1,3,1)%
\mathit{m_{c}}\,\mathit{m_{b}}+10\,I_{1}^{[0,1]}(2,3,1)\mathit{m_{c}}\,%
\mathit{m_{b}} \\
&&+10\,I_{2}^{[0,1]}(2,3,1)\mathit{m_{c}}\,\mathit{m_{b}}+10%
\,I_{1}^{[0,1]}(2,2,2){\mathit{m_{b}}}^{2}-30\,I_{2}^{[0,1]}(1,4,1){\mathit{%
m_{b}}}^{2} \\
&&+15\,I_{{0}}(1,3,1){\mathit{m_{b}}}^{2}+5\,I_{{0}}(2,2,1){\mathit{m_{b}}}%
^{2}+5\,I_{{2}}(2,2,1){\mathit{m_{b}}}^{2} \\
&&-5\,I_{2}^{[0,2]}(3,2,2){\mathit{m_{b}}}^{2}-5\,I_{{0}}(3,1,1){\mathit{%
m_{b}}}^{2}-5\,I_{{0}}(1,2,2){\mathit{m_{b}}}^{2} \\
&&+10\,I_{1}^{[0,1]}(3,2,1){\mathit{m_{b}}}^{2}-5\,I_{1}^{[0,2]}(3,2,2){%
\mathit{m_{b}}}^{2}-30\,I_{0}^{[0,1]}(1,4,1){\mathit{m_{b}}}^{2} \\
&&+10\,I_{2}^{[0,1]}(3,2,1){\mathit{m_{b}}}^{2}+10\,I_{2}^{[0,1]}(2,2,2){%
\mathit{m_{b}}}^{2}-30\,I_{1}^{[0,1]}(1,4,1){\mathit{m_{b}}}^{2} \\
&&-5\,I_{{0}}(2,1,2){\mathit{m_{b}}}^{2}+15\,I_{0}^{[0,1]}(2,2,2){\mathit{%
m_{b}}}^{2}+5\,I_{{1}}(2,2,1){\mathit{m_{b}}}^{2} \\
&&-10I_{1}^{[0,2]}(3,1,2)+10\,I_{{2}}(1,2,1)+5\,I_{{0}}(2,1,1)+10%
\,I_{0}^{[0,1]}(2,1,2) \\
&&+10\,I_{{1}}(1,1,2)+10\,I_{{2}}(1,1,2)+5\,I_{{0}}(1,2,1)+10%
\,I_{0}^{[0,1]}(2,2,1) \\
&&-10\,I_{2}^{[0,2]}(3,2,1)-15\,I_{2}^{[0,2]}(2,2,2)-10%
\,I_{1}^{[0,2]}(3,2,1)+5\,I_{{0}}(1,1,2) \\
&&+5\,I_{1}^{[0,1]}(2,2,1)+5\,I_{2}^{[0,1]}(2,1,2)+5\,I_{1}^{[0,1]}(2,1,2)+5%
\,I_{2}^{[0,1]}(2,2,1) \\
&&-15\,I_{0}^{[0,2]}(2,2,2)-15\,I_{1}^{[0,2]}(2,2,2)+10\,I_{{1}%
}(1,2,1)+10\,I_{{1}}(2,1,1) \\
&&+10\,I_{{2}}(2,1,1)+10\,I_{0}^{[0,1]}(3,1,1)-10\,I_{2}^{[0,2]}(3,1,2)+10%
\,I_{0}^{[0,1]}(1,2,2) \\
&&+10\,I_{1}^{[0,1]}(1,2,2)+10\,I_{2}^{[0,1]}(1,2,2)~~,
\end{eqnarray*}
\begin{eqnarray*}
&&C_{-}=5\,I_{{2}}(3,2,2){\mathit{m_{c}}}^{6}-5\,I_{{1}}(3,2,2){\mathit{m_{c}%
}}^{6}-5\,I_{{0}}(3,2,2){\mathit{m_{c}}}^{5}\mathit{m_{b}} \\
&&+5\,I_{{1}}(3,2,2){\mathit{m_{c}}}^{5}\mathit{m_{b}}-5\,I_{{2}}(3,2,2){%
\mathit{m_{c}}}^{5}\mathit{m_{b}}+5\,I_{{1}}(3,2,2){\mathit{m_{c}}}^{4}{%
\mathit{m_{b}}}^{2} \\
&&-5\,I_{{2}}(3,2,2){\mathit{m_{c}}}^{4}{\mathit{m_{b}}}^{2}-5\,I_{{1}%
}(3,2,2){\mathit{m_{c}}}^{3}{\mathit{m_{b}}}^{3}+5\,I_{{0}}(3,2,2){\mathit{%
m_{c}}}^{3}{\mathit{m_{b}}}^{3} \\
&&+5\,I_{{2}}(3,2,2){\mathit{m_{c}}}^{3}{\mathit{m_{b}}}^{3}-15%
\,I_{2}^{[0,1]}(3,2,2){\mathit{m_{c}}}^{4}-5\,I_{{1}}(3,1,2){\mathit{m_{c}}}%
^{4} \\
&&+15\,I_{{2}}(2,2,2){\mathit{m_{c}}}^{4}-15\,I_{{1}}(4,1,1){\mathit{m_{c}}}%
^{4}+5\,I_{{2}}(3,2,1){\mathit{m_{c}}}^{4} \\
&&+5\,I_{{2}}(3,1,2){\mathit{m_{c}}}^{4}+15\,I_{1}^{[0,1]}(3,2,2){\mathit{%
m_{c}}}^{4}-5\,I_{{1}}(3,2,1){\mathit{m_{c}}}^{4} \\
&&-15\,I_{{1}}(2,2,2){\mathit{m_{c}}}^{4}+15\,I_{{2}}(4,1,1){\mathit{m_{c}}}%
^{4}+5\,I_{{1}}(3,1,2){\mathit{m_{c}}}^{3}\mathit{m_{b}} \\
&&+10\,I_{2}^{[0,1]}(3,2,2){\mathit{m_{c}}}^{3}\mathit{m_{b}}%
+10\,I_{0}^{[0,1]}(3,2,2){\mathit{m_{c}}}^{3}\mathit{m_{b}}-5\,I_{{2}}(3,1,2)%
{\mathit{m_{c}}}^{3}\mathit{m_{b}} \\
&&-10\,I_{{1}}(2,3,1){\mathit{m_{c}}}^{3}\mathit{m_{b}}-15\,I_{{0}}(4,1,1){%
\mathit{m_{c}}}^{3}\mathit{m_{b}}+10\,I_{{2}}(2,3,1){\mathit{m_{c}}}^{3}%
\mathit{m_{b}} \\
&&+15\,I_{{1}}(4,1,1){\mathit{m_{c}}}^{3}\mathit{m_{b}}-15\,I_{{2}}(4,1,1){%
\mathit{m_{c}}}^{3}\mathit{m_{b}}-10\,I_{{0}}(2,3,1){\mathit{m_{c}}}^{3}%
\mathit{m_{b}} \\
&&+5\,I_{{0}}(3,2,1){\mathit{m_{c}}}^{3}\mathit{m_{b}}-5\,I_{{0}}(3,1,2){%
\mathit{m_{c}}}^{3}\mathit{m_{b}}-10\,I_{{2}}(2,2,2){\mathit{m_{c}}}^{3}%
\mathit{m_{b}} \\
&&-10\,I_{{0}}(2,2,2){\mathit{m_{c}}}^{3}\mathit{m_{b}}-10%
\,I_{1}^{[0,1]}(3,2,2){\mathit{m_{c}}}^{3}\mathit{m_{b}}+5\,I_{{1}}(3,2,1){%
\mathit{m_{c}}}^{3}\mathit{m_{b}} \\
&&-5\,I_{{2}}(3,2,1){\mathit{m_{c}}}^{3}\mathit{m_{b}}+10\,I_{{1}}(2,2,2){%
\mathit{m_{c}}}^{3}\mathit{m_{b}}-5\,I_{0}^{[0,1]}(3,2,2){\mathit{m_{c}}}^{2}%
{\mathit{m_{b}}}^{2} \\
&&+5\,I_{{0}}(2,2,2){\mathit{m_{c}}}^{2}{\mathit{m_{b}}}^{2}+30\,I_{{1}%
}(1,4,1){\mathit{m_{c}}}^{2}{\mathit{m_{b}}}^{2}+10\,I_{{2}}(3,2,1){\mathit{%
m_{c}}}^{2}{\mathit{m_{b}}}^{2} \\
&&-10\,I_{{1}}(3,2,1){\mathit{m_{c}}}^{2}{\mathit{m_{b}}}^{2}-30\,I_{{2}%
}(1,4,1){\mathit{m_{c}}}^{2}{\mathit{m_{b}}}^{2}+5\,I_{0}^{[0,1]}(3,2,2)%
\mathit{m_{c}}\,{\mathit{m_{b}}}^{3} \\
&&+30\,I_{{0}}(1,4,1)\mathit{m_{c}}\,{\mathit{m_{b}}}^{3}-30\,I_{{1}}(1,4,1)%
\mathit{m_{c}}\,{\mathit{m_{b}}}^{3}+10\,I_{{1}}(2,3,1)\mathit{m_{c}}\,{%
\mathit{m_{b}}}^{3} \\
&&+10\,I_{{1}}(3,2,1)\mathit{m_{c}}\,{\mathit{m_{b}}}^{3}-5%
\,I_{1}^{[0,1]}(3,2,2)\mathit{m_{c}}\,{\mathit{m_{b}}}^{3}+30\,I_{{2}}(1,4,1)%
\mathit{m_{c}}\,{\mathit{m_{b}}}^{3} \\
&&-10\,I_{{2}}(2,3,1)\mathit{m_{c}}\,{\mathit{m_{b}}}^{3}-10\,I_{{2}}(3,2,1)%
\mathit{m_{c}}\,{\mathit{m_{b}}}^{3}+5\,I_{2}^{[0,1]}(3,2,2)\mathit{m_{c}}\,{%
\mathit{m_{b}}}^{3} \\
&&+5\,I_{{0}}(3,2,1){\mathit{m_{b}}}^{4}-15\,I_{{0}}(1,4,1){\mathit{m_{b}}}%
^{4}+15\,I_{1}^{[0,1]}(4,1,1){\mathit{m_{c}}}^{2} \\
&&-10\,I_{{0}}(2,2,1){\mathit{m_{c}}}^{2}+10\,I_{0}^{[0,1]}(3,2,1){\mathit{%
m_{c}}}^{2}+10\,I_{{0}}(2,1,2){\mathit{m_{c}}}^{2} \\
&&+30\,I_{1}^{[0,1]}(2,2,2){\mathit{m_{c}}}^{2}+10\,I_{{2}}(1,2,2){\mathit{%
m_{c}}}^{2}-10\,I_{{1}}(1,2,2){\mathit{m_{c}}}^{2} \\
&&-30\,I_{2}^{[0,1]}(2,2,2){\mathit{m_{c}}}^{2}-15\,I_{2}^{[0,1]}(3,1,2){%
\mathit{m_{c}}}^{2}-15\,I_{2}^{[0,1]}(4,1,1){\mathit{m_{c}}}^{2} \\
&&+15\,I_{1}^{[0,1]}(3,1,2){\mathit{m_{c}}}^{2}-15\,I_{2}^{[0,1]}(3,2,1){%
\mathit{m_{c}}}^{2}+15\,I_{1}^{[0,1]}(3,2,1){\mathit{m_{c}}}^{2} \\
&&-10\,I_{0}^{[0,1]}(3,1,2){\mathit{m_{c}}}^{2}-15\,I_{1}^{[0,2]}(3,2,2){%
\mathit{m_{c}}}^{2}+15\,I_{2}^{[0,2]}(3,2,2){\mathit{m_{c}}}^{2} \\
&&+40\,I_{{2}}(1,3,1)\mathit{m_{c}}\,\mathit{m_{b}}-5\,I_{0}^{[0,2]}(3,2,2)%
\mathit{m_{c}}\,\mathit{m_{b}}-5\,I_{1}^{[0,1]}(3,2,1)\mathit{m_{c}}\,%
\mathit{m_{b}} \\
&&+15\,I_{0}^{[0,1]}(3,1,2)\mathit{m_{c}}\,\mathit{m_{b}}-10\,I_{{1}}(1,2,2)%
\mathit{m_{c}}\,\mathit{m_{b}}+15\,I_{2}^{[0,1]}(3,1,2)\mathit{m_{c}}\,%
\mathit{m_{b}} \\
&&-15\,I_{1}^{[0,1]}(3,1,2)\mathit{m_{c}}\,\mathit{m_{b}}-10%
\,I_{1}^{[0,1]}(2,2,2)\mathit{m_{c}}\,\mathit{m_{b}}+5\,I_{2}^{[0,1]}(3,2,1)%
\mathit{m_{c}}\,\mathit{m_{b}} \\
&&+20\,I_{{0}}(1,3,1)\mathit{m_{c}}\,\mathit{m_{b}}+5\,I_{1}^{[0,2]}(3,2,2)%
\mathit{m_{c}}\,\mathit{m_{b}}+5\,I_{{2}}(2,1,2)\mathit{m_{c}}\,\mathit{m_{b}%
} \\
&&+20\,I_{{0}}(2,2,1)\mathit{m_{c}}\,\mathit{m_{b}}-20\,I_{{1}}(2,2,1)%
\mathit{m_{c}}\,\mathit{m_{b}}+10\,I_{{2}}(1,2,2)\mathit{m_{c}}\,\mathit{%
m_{b}} \\
&&+10\,I_{0}^{[0,1]}(2,2,2)\mathit{m_{c}}\,\mathit{m_{b}}-10%
\,I_{2}^{[0,1]}(2,3,1)\mathit{m_{c}}\,\mathit{m_{b}}+15\,I_{0}^{[0,1]}(3,2,1)%
\mathit{m_{c}}\,\mathit{m_{b}} \\
&&-5\,I_{{1}}(2,1,2)\mathit{m_{c}}\,\mathit{m_{b}}+10\,I_{2}^{[0,1]}(2,2,2)%
\mathit{m_{c}}\,\mathit{m_{b}}+10\,I_{1}^{[0,1]}(2,3,1)\mathit{m_{c}}\,%
\mathit{m_{b}} \\
&&+10\,I_{{0}}(1,2,2)\mathit{m_{c}}\,\mathit{m_{b}}+5\,I_{{0}}(2,1,2)\mathit{%
m_{c}}\,\mathit{m_{b}}-5\,I_{2}^{[0,2]}(3,2,2)\mathit{m_{c}}\,\mathit{m_{b}}
\\
&&+20\,I_{{2}}(2,2,1)\mathit{m_{c}}\,\mathit{m_{b}}-40\,I_{{1}}(1,3,1)%
\mathit{m_{c}}\,\mathit{m_{b}}+10\,I_{0}^{[0,1]}(2,3,1)\mathit{m_{c}}\,%
\mathit{m_{b}} \\
&&-15\,I_{{0}}(1,3,1){\mathit{m_{b}}}^{2}-30\,I_{1}^{[0,1]}(1,4,1){\mathit{%
m_{b}}}^{2}+5\,I_{{0}}(2,1,2){\mathit{m_{b}}}^{2} \\
&&-10\,I_{2}^{[0,1]}(3,2,1){\mathit{m_{b}}}^{2}+10\,I_{1}^{[0,1]}(3,2,1){%
\mathit{m_{b}}}^{2}-5\,I_{{2}}(2,2,1){\mathit{m_{b}}}^{2} \\
&&+5\,I_{{1}}(2,2,1){\mathit{m_{b}}}^{2}-5\,I_{1}^{[0,2]}(3,2,2){\mathit{%
m_{b}}}^{2}+5\,I_{{0}}(3,1,1){\mathit{m_{b}}}^{2} \\
&&+5\,I_{2}^{[0,2]}(3,2,2){\mathit{m_{b}}}^{2}+30\,I_{2}^{[0,1]}(1,4,1){%
\mathit{m_{b}}}^{2}-5\,I_{0}^{[0,1]}(2,2,2){\mathit{m_{b}}}^{2} \\
&&+10\,I_{1}^{[0,1]}(2,2,2){\mathit{m_{b}}}^{2}-10\,I_{{0}}(2,2,1){\mathit{%
m_{b}}}^{2}-10\,I_{2}^{[0,1]}(2,2,2){\mathit{m_{b}}}^{2} \\
&&-10\,I_{{2}}(2,1,1)+10\,I_{{1}}(1,1,2)-10\,I_{{2}}(1,1,2)+10%
\,I_{1}^{[0,1]}(1,2,2) \\
&&-10\,I_{{2}}(1,2,1)+5\,I_{0}^{[0,1]}(2,2,1)+10\,I_{2}^{[0,2]}(3,2,1)-15%
\,I_{1}^{[0,2]}(2,2,2) \\
&&+15\,I_{2}^{[0,2]}(2,2,2)+5\,I_{1}^{[0,1]}(2,1,2)-10%
\,I_{1}^{[0,2]}(3,2,1)-10\,I_{1}^{[0,2]}(3,1,2) \\
&&+10\,I_{2}^{[0,2]}(3,1,2)-5\,I_{0}^{[0,1]}(2,1,2)-10%
\,I_{2}^{[0,1]}(1,2,2)+5\,I_{1}^{[0,1]}(2,2,1) \\
&&-5\,I_{2}^{[0,1]}(2,1,2)-5\,I_{2}^{[0,1]}(2,2,1)+10\,I_{{1}}(1,2,1)+10\,I_{%
{1}}(2,1,1)~.
\end{eqnarray*}
\begin{eqnarray*}
&&C_{T}=-5\,I_{{0}}(3,2,2){{m_{c}}}^{4}{{m_{b}}}^{3}+5\,I_{{0}}(3,2,2){{m_{c}}}%
^{2}{{m_{b}}}^{5}-5\,I_{{0}}(3,1,2){{m_{c}}}^{4}{m_{b}} \\
&&+5\,I_{{0}}(3,2,1){{m_{c}}}^{4}{m_{b}}-5\,I_{{0}}(3,1,2){{m_{c}}}^{3}{{%
m_{b}}}^{2}+5\,I_{{0}}(3,2,1){{m_{c}}}^{3}{{m_{b}}}^{2} \\
&&+10\,I_{0}^{[0,1]}(3,2,2){{m_{c}}}^{2}{{m_{b}}}^{3}-15\,I_{{0}}(3,2,1){{%
m_{c}}}^{2}{{m_{b}}}^{3}-15\,I_{{0}}(4,1,1){{m_{c}}}^{2}{{m_{b}}}^{3} \\
&&-10\,I_{{0}}(2,3,1){{m_{c}}}^{2}{{m_{b}}}^{3}+5\,I_{{0}}(3,1,2){{m_{c}}}%
^{2}{{m_{b}}}^{3}-10\,I_{{0}}(2,2,2){{m_{c}}}^{2}{{m_{b}}}^{3} \\
&&+5\,I_{0}^{[0,1]}(3,2,2){{m_{b}}}^{5}-5\,I_{{0}}(2,2,2){{m_{b}}}%
^{5}-10\,I_{{0}}(2,3,1){{m_{b}}}^{5} \\
&&-10\,I_{{0}}(3,2,1){{m_{b}}}^{5}+30\,I_{{0}}(1,4,1){{m_{b}}}^{5}+10\,I_{{0}%
}(3,1,1){{m_{c}}}^{3} \\
&&-10\,I_{{0}}(2,1,2){{m_{c}}}^{2}{m_{b}}-10\,I_{{0}}(3,1,1){{m_{c}}}^{2}{%
m_{b}}-10\,I_{0}^{[0,1]}(3,2,1){{m_{c}}}^{2}{m_{b}} \\
&&+10\,I_{0}^{[0,1]}(3,1,2){{m_{c}}}^{2}{m_{b}}-5\,I_{0}^{[0,1]}(3,2,1){%
m_{c}}\,{{m_{b}}}^{2}+5\,I_{{0}}(2,2,1){m_{c}}\,{{m_{b}}}^{2} \\
&&-5\,I_{{0}}(2,1,2){m_{c}}\,{{m_{b}}}^{2}+5\,I_{0}^{[0,1]}(3,1,2){m_{c}}\,%
{{m_{b}}}^{2}-10\,I_{{0}}(1,2,2){{m_{b}}}^{3} \\
&&-5\,I_{{0}}(2,2,1){{m_{b}}}^{3}-5\,I_{0}^{[0,2]}(3,2,2){{m_{b}}}%
^{3}+50\,I_{{0}}(1,3,1){{m_{b}}}^{3} \\
&&+15\,I_{0}^{[0,1]}(3,1,2){{m_{b}}}^{3}+5\,I_{0}^{[0,1]}(3,2,1){{m_{b}}}%
^{3}-10\,I_{{0}}(3,1,1){{m_{b}}}^{3} \\
&&-10\,I_{{0}}(2,1,2){{m_{b}}}^{3}+10\,I_{0}^{[0,1]}(2,3,1){{m_{b}}}%
^{3}+10\,I_{0}^{[0,1]}(2,2,2){{m_{b}}}^{3} \\
&&-10\,I_{0}^{[0,1]}(3,1,1){m_{c}}+10\,I_{{0}}(2,1,1){m_{c}}+30\,I_{{0}%
}(1,2,1){m_{b}} \\
&&-10\,I_{{0}}(1,1,2)\mathit{m_{b}}+5\,I_{{0}}(2,1,1){m_{b}}%
-5\,I_{0}^{[0,2]}(3,1,2){m_{b}} \\
&&+5\,I_{0}^{[0,2]}(3,2,1){m_{b}}+10\,I_{0}^{[0,1]}(2,1,2){m_{b}}
\end{eqnarray*}
where
\begin{eqnarray}
\hat{I}_n^{[i,j]} (a,b,c) = \left( M_1^2 \right)^i \left( M_2^2 \right)^j
\frac{d^i}{d\left( M_1^2 \right)^i} \frac{d^j}{d\left( M_2^2 \right)^j} %
\left[\left( M_1^2 \right)^i \left( M_2^2 \right)^j \hat{I}_n(a,b,c) \right]%
~.  \nonumber
\end{eqnarray}
\begin{center}
{\Large \textbf{Appendix--B}}
\end{center}


\setcounter{equation}{0}
\renewcommand{\theequation}{C.\arabic{equation}}
\setcounter{section}{0} \setcounter{table}{0}

In this appendix,  the explicit expressions of the coefficients of
the gluon condensate  entering  the HQET limit of the  form factors
$\tilde{f}_{+}^{HQET},~\tilde{f}_{-}^{HQET}$ and
$\tilde{f}_{T}^{HQET}$ are given. Note that only in this appendix,
by $\hat{I}_{i}(a,b,c)$ we mean $\hat{I}_{i}(a,b,c)^{HQET}$which are
defined in Eq. (\ref{e7326guy}).
\begin{eqnarray*}
C^{HQET}_{+} &=&2\,{\frac{\hat{I}_{2}^{[0,1]}(3,2,2){{m_{b}}}^{5}}{\sqrt{z}}}+{\frac{%
\hat{I}_{0}^{[0,1]}(3,2,2){{m_{b}}}^{4}}{\sqrt{z}}}-32\,{\frac{\hat{I}_{{2}}(2,1,1){{%
m_{b}}}^{4}}{\sqrt{z}}} \\
&&-4\,{\frac{\hat{I}_{{0}}(2,2,1){{m_{b}}}^{4}}{\sqrt{z}}}-8\,{\frac{\hat{I}_{{0}}(2,1,1)%
{{m_{b}}}^{3}}{\sqrt{z}}}+16\,{\frac{\hat{I}_{2}^{[0,2]}(3,1,2){{m_{b}}}^{3}}{%
\sqrt{z}}} \\
&&-8\,{\frac{\hat{I}_{0}^{[0,1]}(3,1,1){{m_{b}}}^{2}}{\sqrt{z}}}+4\,{\frac{\hat{I}_{{0}%
}(2,1,2){{m_{b}}}^{4}}{\sqrt{z}}}+12\,{\frac{\hat{I}_{2}^{[0,1]}(3,1,2){{m_{b}}}%
^{4}}{\sqrt{z}}} \\
&&-16\,{\frac{\hat{I}_{{0}}(1,1,2){{m_{b}}}^{4}}{\sqrt{z}}}-12\,{\frac{%
\hat{I}_{0}^{[0,1]}(2,2,2){{m_{b}}}^{4}}{\sqrt{z}}}-8\,{\frac{\hat{I}_{2}^{[0,1]}(3,2,1){%
{m_{b}}}^{4}}{\sqrt{z}}} \\
&&+8\,{\frac{\hat{I}_{{0}}(1,2,2){{m_{b}}}^{5}}{\sqrt{z}}}+8\,{\frac{\hat{I}_{{2}}(2,1,2)%
{{m_{b}}}^{5}}{\sqrt{z}}}-16\,{\frac{\hat{I}_{2}^{[0,1]}(2,1,2){{m_{b}}}^{4}}{%
\sqrt{z}}} \\
&&+6\,{\frac{\hat{I}_{0}^{[0,1]}(3,1,2){{m_{b}}}^{3}}{\sqrt{z}}}-16\,{\frac{%
\hat{I}_{0}^{[0,1]}(2,1,2){{m_{b}}}^{3}}{\sqrt{z}}}-64\,{\frac{\hat{I}_{{2}}(1,1,2){{%
m_{b}}}^{5}}{\sqrt{z}}} \\
&&+4\,{\frac{\hat{I}_{2}^{[0,2]}(3,2,2){{m_{b}}}^{4}}{\sqrt{z}}}-4\,{\frac{\hat{I}_{{2}%
}(3,2,1){{m_{b}}}^{5}}{\sqrt{z}}}-8\,{\frac{\hat{I}_{{2}}(2,2,1){{m_{b}}}^{5}}{%
\sqrt{z}}} \\
&&-16\,{\frac{\hat{I}_{2}^{[0,1]}(2,2,2){{m_{b}}}^{5}}{\sqrt{z}}}+16\,{\frac{%
\hat{I}_{1}^{[0,2]}(3,1,2){{m_{b}}}^{3}}{z}}-2\,{\frac{\hat{I}_{0}^{[{0,2]}}(3,2,2){{%
m_{b}}}^{3}}{z}} \\
&&+24\,{\frac{\hat{I}_{0}^{[0,2]}(2,2,2){{m_{b}}}^{3}}{z}}-6\,{\frac{%
\hat{I}_{2}^{[0,1]}(4,1,1){{m_{b}}}^{3}}{z}}+4\,{\frac{\hat{I}_{0}^{[0,1]}(3,2,1){{m_{b}}%
}^{3}}{z}} \\
&&-64\,{\frac{\hat{I}_{2}^{[0,1]}(1,2,2){{m_{b}}}^{5}}{z}}-16\,{\frac{%
\hat{I}_{0}^{[0,1]}(2,2,1){{m_{b}}}^{3}}{z}}-64\,{\frac{\hat{I}_{{1}}(1,1,2){{m_{b}}}^{5}%
}{z}} \\
&&+2\,{\frac{\hat{I}_{{0}}(3,1,1){{m_{b}}}^{3}}{z}}-8\,{\frac{\hat{I}_{{1}}(2,2,1){{m_{b}%
}}^{5}}{z}}-3/2\,{\frac{\hat{I}_{{0}}(4,1,1){{m_{b}}}^{3}}{z}} \\
&&+16\,{\frac{\hat{I}_{2}^{[0,2]}(3,2,1){{m_{b}}}^{3}}{z}}+2\,{\frac{\hat{I}_{{0}}(2,2,2)%
{{m_{b}}}^{5}}{z}}+1/2\,{\frac{\hat{I}_{{0}}(3,2,2){{m_{b}}}^{5}}{z}} \\
&&-24\,{\frac{\hat{I}_{{0}}(1,3,1){{m_{b}}}^{5}}{z}}+{\frac{\hat{I}_{{2}}(3,2,2){{m_{b}}}%
^{6}}{z}}+16\,{\frac{\hat{I}_{{0}}(1,2,2){{m_{b}}}^{5}}{z}} \\
&&-8\,{\frac{\hat{I}_{0}^{[0,1]}(3,1,2){{m_{b}}}^{3}}{z}}-3\,{\frac{%
\hat{I}_{0}^{[0,1]}(4,1,1){{m_{b}}}^{2}}{z}}+2\,{\frac{\hat{I}_{1}^{[0,1]}(3,2,2){{m_{b}}%
}^{5}}{z}} \\
&&-4\,{\frac{\hat{I}_{{1}}(3,2,1){{m_{b}}}^{5}}{z}}-2\,{\frac{\hat{I}_{{2}}(3,1,2){{m_{b}%
}}^{5}}{z}}-16\,{\frac{\hat{I}_{2}^{[0,1]}(2,2,1){{m_{b}}}^{4}}{z}} \\
&&+8\,{\frac{\hat{I}_{{1}}(2,1,2){{m_{b}}}^{5}}{z}}+4\,{\frac{\hat{I}_{{2}}(3,2,1){{m_{b}%
}}^{5}}{z}}-4\,{\frac{\hat{I}_{2}^{[0,2]}(3,2,2){{m_{b}}}^{4}}{z}} \\
&&+12\,{\frac{\hat{I}_{0}^{[{0,1]}}(3,1,2){{m_{b}}}^{4}}{z}}-32\,{\frac{\hat{I}_{{1}%
}(2,1,1){{m_{b}}}^{4}}{z}}+16\,{\frac{\hat{I}_{2}^{[0,1]}(2,2,2){{m_{b}}}^{5}}{z}}
\\
&&-12\,{\frac{\hat{I}_{2}^{[0,1]}(3,1,2){{m_{b}}}^{4}}{z}}+32\,{\frac{\hat{I}_{{2}%
}(2,2,1){{m_{b}}}^{5}}{z}}-{\frac{\hat{I}_{{0}}(3,1,2){{m_{b}}}^{4}}{z}} \\
&&+20\,{\frac{\hat{I}_{{0}}(2,2,1){{m_{b}}}^{4}}{z}}-8\,{\frac{\hat{I}_{1}^{[0,1]}(3,2,1)%
{{m_{b}}}^{4}}{z}}+8\,{\frac{\hat{I}_{0}^{[{0,1]}}(2,2,2){{m_{b}}}^{4}}{z}} \\
&&-3\,{\frac{\hat{I}_{{2}}(4,1,1){{m_{b}}}^{4}}{z}}-{\frac{\hat{I}_{0}^{[{0,1]}}(3,2,2){{%
m_{b}}}^{4}}{z}}+48\,{\frac{\hat{I}_{2}^{[0,2]}(2,2,2){{m_{b}}}^{4}}{z}} \\
&&-8\,{\frac{\hat{I}_{{2}}(2,3,1){{m_{b}}}^{6}}{z}}-16\,{\frac{\hat{I}_{{0}}(1,2,1){{%
m_{b}}}^{4}}{z}}-16\,{\frac{\hat{I}_{1}^{[0,1]}(2,2,2){{m_{b}}}^{5}}{z}} \\
&&+4\,{\frac{\hat{I}_{{0}}(2,1,2){{m_{b}}}^{4}}{z}}-64\,{\frac{\hat{I}_{{2}}(1,2,1){{%
m_{b}}}^{5}}{z}}-32\,{\frac{\hat{I}_{0}^{[0,1]}(1,2,2){{m_{b}}}^{4}}{z}} \\
&&-16\,{\frac{\hat{I}_{0}^{[{0,1]}}(2,1,2){{m_{b}}}^{4}}{z}}+32\,{\frac{\hat{I}_{{2}%
}(1,2,2){{m_{b}}}^{6}}{z}}+4\,{\frac{\hat{I}_{1}^{[0,2]}(3,2,2){{m_{b}}}^{4}}{z}}
\\
&&-12\,{\frac{\hat{I}_{{0}}(1,4,1){{m_{b}}}^{6}}{z}}+4\,{\frac{\hat{I}_{2}^{[0,1]}(3,2,1)%
{{m_{b}}}^{4}}{z}}+3\,{\frac{\hat{I}_{{2}}(4,1,1){{m_{b}}}^{4}}{{z}^{3/2}}} \\
&&+16\,{\frac{\hat{I}_{1}^{[0,1]}(2,2,2){{m_{b}}}^{5}}{{z}^{3/2}}}-3\,{\frac{\hat{I}_{{1}%
}(4,1,1){{m_{b}}}^{4}}{{z}^{3/2}}}+4\,{\frac{\hat{I}_{{1}}(3,2,1){{m_{b}}}^{5}}{{z}%
^{3/2}}} \\
&&-12\,{\frac{\hat{I}_{1}^{[0,1]}(3,1,2){{m_{b}}}^{4}}{{z}^{3/2}}}-16\,{\frac{%
\hat{I}_{1}^{[0,1]}(2,2,1){{m_{b}}}^{4}}{{z}^{3/2}}}-4\,{\frac{\hat{I}_{{0}}(2,2,2){{%
m_{b}}}^{5}}{{z}^{3/2}}} \\
&&+6\,{\frac{\hat{I}_{0}^{[0,2]}(3,2,2){{m_{b}}}^{3}}{{z}^{3/2}}}+48\,{\frac{%
\hat{I}_{0}^{[0,1]}(1,4,1){{m_{b}}}^{5}}{{z}^{3/2}}}+16\,{\frac{%
\hat{I}_{1}^{[0,2]}(3,2,1){{m_{b}}}^{3}}{{z}^{3/2}}} \\
&&-64\,{\frac{\hat{I}_{{1}}(1,2,1){{m_{b}}}^{5}}{{z}^{3/2}}}-8\,{\frac{%
\hat{I}_{0}^{[0,1]}(2,3,1){{m_{b}}}^{4}}{{z}^{3/2}}}-4\,{\frac{\hat{I}_{1}^{[0,2]}(3,2,2)%
{{m_{b}}}^{4}}{{z}^{3/2}}} \\
&&+4\,{\frac{\hat{I}_{1}^{[0,1]}(3,2,1){{m_{b}}}^{4}}{{z}^{3/2}}}+3/2\,{\frac{\hat{I}_{{0%
}}(4,1,1){{m_{b}}}^{3}}{{z}^{3/2}}}+2\,{\frac{\hat{I}_{{2}}(3,1,2){{m_{b}}}^{5}}{{z%
}^{3/2}}} \\
&&+12\,{\frac{\hat{I}_{2}^{[0,2]}(3,2,2){{m_{b}}}^{4}}{{z}^{3/2}}}-48\,{\frac{%
\hat{I}_{2}^{[0,1]}(2,2,2){{m_{b}}}^{5}}{{z}^{3/2}}}-64\,{\frac{%
\hat{I}_{1}^{[0,1]}(1,2,2){{m_{b}}}^{5}}{{z}^{3/2}}}\\
&&-64\,{\frac{\hat{I}_{1}^{[0,1]}(1,2,2){{m_{b}}}^{5}}{{z}^{3/2}}}-2\,{\frac{\hat{I}_{{0}%
}(3,2,1){{m_{b}}}^{4}}{{z}^{3/2}}}+4\,{\frac{\hat{I}_{{0}}(2,2,1){{m_{b}}}^{4}}{{z}%
^{3/2}}} \\
&&-2\,{\frac{\hat{I}_{{2}}(3,2,1){{m_{b}}}^{5}}{{z}^{3/2}}}+16\,{\frac{\hat{I}_{{0}%
}(1,2,2){{m_{b}}}^{5}}{{z}^{3/2}}}+32\,{\frac{\hat{I}_{{1}}(1,2,2){{m_{b}}}^{6}}{{z%
}^{3/2}}} \\
&&-8\,{\frac{\hat{I}_{{1}}(2,3,1){{m_{b}}}^{6}}{{z}^{3/2}}}+32\,{\frac{\hat{I}_{{2}%
}(1,2,2){{m_{b}}}^{6}}{{z}^{3/2}}}-6\,{\frac{\hat{I}_{1}^{[0,1]}(4,1,1){{m_{b}}}%
^{3}}{{z}^{3/2}}} \\
&&+128\,{\frac{\hat{I}_{{2}}(1,3,1){{m_{b}}}^{6}}{{z}^{3/2}}}+32\,{\frac{\hat{I}_{{1}%
}(2,2,1){{m_{b}}}^{5}}{{z}^{3/2}}}+4\,{\frac{\hat{I}_{2}^{[0,1]}(3,2,2){{m_{b}}}%
^{5}}{{z}^{3/2}}} \\
&&-24\,{\frac{\hat{I}_{0}^{[0,1]}(2,2,2){{m_{b}}}^{4}}{{z}^{3/2}}}-8\,{\frac{%
\hat{I}_{0}^{[0,1]}(3,2,1){{m_{b}}}^{3}}{{z}^{3/2}}}+64\,{\frac{\hat{I}_{{0}}(1,3,1){{%
m_{b}}}^{5}}{{z}^{3/2}}} \\
&&-2\,{\frac{\hat{I}_{{1}}(3,1,2){{m_{b}}}^{5}}{{z}^{3/2}}}-16\,{\frac{%
\hat{I}_{2}^{[0,1]}(2,3,1){{m_{b}}}^{5}}{{z}^{3/2}}}-8\,{\frac{\hat{I}_{{2}}(2,2,2){{%
m_{b}}}^{6}}{{z}^{3/2}}} \\
&&+24\,{\frac{\hat{I}_{{0}}(1,4,1){{m_{b}}}^{6}}{{z}^{3/2}}}+{\frac{\hat{I}_{{1}}(3,2,2){%
{m_{b}}}^{6}}{{z}^{3/2}}}+48\,{\frac{\hat{I}_{1}^{[0,2]}(2,2,2){{m_{b}}}^{4}}{{z}%
^{3/2}}} \\
&&-{\frac{\hat{I}_{{2}}(3,2,2){{m_{b}}}^{6}}{{z}^{3/2}}}+48\,{\frac{\hat{I}_{{2}}(1,4,1){%
{m_{b}}}^{7}}{{z}^{3/2}}}+96\,{\frac{\hat{I}_{2}^{[0,1]}(1,4,1){{m_{b}}}^{6}}{{z}%
^{3/2}}} \\
&&+2\,{\frac{\hat{I}_{0}^{[0,1]}(3,2,2){{m_{b}}}^{4}}{{z}^{3/2}}}-12\,{\frac{%
\hat{I}_{2}^{[0,1]}(3,2,1){{m_{b}}}^{4}}{{z}^{3/2}}}+6\,{\frac{\hat{I}_{{0}}(2,2,2){{%
m_{b}}}^{5}}{{z}^{2}}} \\
&&-2\,{\frac{\hat{I}_{{1}}(3,2,1){{m_{b}}}^{5}}{{z}^{2}}}-48\,{\frac{%
\hat{I}_{1}^{[0,1]}(2,2,2){{m_{b}}}^{5}}{{z}^{2}}}-12\,{\frac{\hat{I}_{1}^{[0,1]}(3,2,1){%
{m_{b}}}^{4}}{{z}^{2}}} \\
&&+2\,{\frac{\hat{I}_{{1}}(3,1,2){{m_{b}}}^{5}}{{z}^{2}}}-{\frac{\hat{I}_{{2}}(3,2,2){{%
m_{b}}}^{6}}{{z}^{2}}}+4\,{\frac{\hat{I}_{{0}}(2,3,1){{m_{b}}}^{5}}{{z}^{2}}} \\
&&+2\,{\frac{\hat{I}_{{2}}(3,2,1){{m_{b}}}^{5}}{{z}^{2}}}-1/2\,{\frac{\hat{I}_{{0}%
}(3,2,2){{m_{b}}}^{5}}{{z}^{2}}}+48\,{\frac{\hat{I}_{{1}}(1,4,1){{m_{b}}}^{7}}{{z}%
^{2}}} \\
&&-3\,{\frac{\hat{I}_{0}^{[0,1]}(3,2,2){{m_{b}}}^{4}}{{z}^{2}}}+3\,{\frac{\hat{I}_{{1}%
}(4,1,1){{m_{b}}}^{4}}{{z}^{2}}}-48\,{\frac{\hat{I}_{{2}}(1,4,1){{m_{b}}}^{7}}{{z}%
^{2}}} \\
&&+12\,{\frac{\hat{I}_{1}^{[0,2]}(3,2,2){{m_{b}}}^{4}}{{z}^{2}}}-6\,{\frac{%
\hat{I}_{2}^{[0,1]}(3,2,2){{m_{b}}}^{5}}{{z}^{2}}}+12\,{\frac{\hat{I}_{{2}}(2,2,2){{m_{b}%
}}^{6}}{{z}^{2}}} \\
&&-24\,{\frac{\hat{I}_{{0}}(1,4,1){{m_{b}}}^{6}}{{z}^{2}}}-8\,{\frac{\hat{I}_{{1}}(2,2,2)%
{{m_{b}}}^{6}}{{z}^{2}}}+8\,{\frac{\hat{I}_{{2}}(2,3,1){{m_{b}}}^{6}}{{z}^{2}}} \\
&&+96\,{\frac{\hat{I}_{1}^{[0,1]}(1,4,1){{m_{b}}}^{6}}{{z}^{2}}}+4\,{\frac{%
\hat{I}_{1}^{[0,1]}(3,2,2){{m_{b}}}^{5}}{{z}^{2}}}-16\,{\frac{\hat{I}_{1}^{[0,1]}(2,3,1){%
{m_{b}}}^{5}}{{z}^{2}}} \\
&&-{\frac{\hat{I}_{{1}}(3,2,2){{m_{b}}}^{6}}{{z}^{2}}}+128\,{\frac{\hat{I}_{{1}}(1,3,1){{%
m_{b}}}^{6}}{{z}^{2}}}+32\,{\frac{\hat{I}_{{1}}(1,2,2){{m_{b}}}^{6}}{{z}^{2}}} \\
&&+2\,{\frac{\hat{I}_{{1}}(3,2,1){{m_{b}}}^{5}}{{z}^{5/2}}}-48\,{\frac{\hat{I}_{{1}%
}(1,4,1){{m_{b}}}^{7}}{{z}^{5/2}}}+12\,{\frac{\hat{I}_{{1}}(2,2,2){{m_{b}}}^{6}}{{z%
}^{5/2}}} \\
&&+1/2\,{\frac{\hat{I}_{{0}}(3,2,2){{m_{b}}}^{5}}{{z}^{5/2}}}-6\,{\frac{%
\hat{I}_{1}^{[0,1]}(3,2,2){{m_{b}}}^{5}}{{z}^{5/2}}}+{\frac{\hat{I}_{{2}}(3,2,2){{m_{b}}}%
^{6}}{{z}^{5/2}}} \\
&&-{\frac{\hat{I}_{{1}}(3,2,2){{m_{b}}}^{6}}{{z}^{5/2}}}+8\,{\frac{\hat{I}_{{1}}(2,3,1){{%
m_{b}}}^{6}}{{z}^{5/2}}}+{\frac{\hat{I}_{{1}}(3,2,2){{m_{b}}}^{6}}{{z}^{3}}} \\
&&+4\,\hat{I}_{{0}}(2,1,2){{m_{b}}}^{4}+2\,\hat{I}_{{0}}(3,1,1){{m_{b}}}^{3}+\hat{I}_{{0}%
}(3,2,1){{m_{b}}}^{4}
\end{eqnarray*}
\begin{eqnarray*}
C^{HQET}_{-} &=&-8\,{\frac{\hat{I}_{{2}}(2,1,2){{m_{b}}}^{5}}{\sqrt{z}}}+32\,{\frac{\hat{I}_{{2}%
}(2,1,1){{m_{b}}}^{4}}{\sqrt{z}}}-4\,{\frac{\hat{I}_{{0}}(2,1,2){{m_{b}}}^{4}}{%
\sqrt{z}}} \\
&&+4\,{\frac{\hat{I}_{0}^{[0,1]}(2,2,2){{m_{b}}}^{4}}{\sqrt{z}}}-12\,{\frac{%
\hat{I}_{2}^{[0,1]}(3,1,2){{m_{b}}}^{4}}{\sqrt{z}}}+64\,{\frac{\hat{I}_{{2}}(1,1,2){{%
m_{b}}}^{5}}{\sqrt{z}}} \\
&&-16\,{\frac{\hat{I}_{2}^{[0,2]}(3,1,2){{m_{b}}}^{3}}{\sqrt{z}}}-6\,{\frac{%
\hat{I}_{0}^{[0,1]}(3,1,2){{m_{b}}}^{3}}{\sqrt{z}}}+8\,{\frac{\hat{I}_{0}^{[0,1]}(2,1,2){%
{m_{b}}}^{3}}{\sqrt{z}}} \\
&&+8\,{\frac{\hat{I}_{2}^{[0,1]}(3,2,1){{m_{b}}}^{4}}{\sqrt{z}}}+16\,{\frac{%
\hat{I}_{2}^{[0,1]}(2,2,2){{m_{b}}}^{5}}{\sqrt{z}}}+8\,{\frac{\hat{I}_{{0}}(2,2,1){{m_{b}%
}}^{4}}{\sqrt{z}}} \\
&&+16\,{\frac{\hat{I}_{2}^{[0,1]}(2,1,2){{m_{b}}}^{4}}{\sqrt{z}}}-2\,{\frac{%
\hat{I}_{2}^{[0,1]}(3,2,2){{m_{b}}}^{5}}{\sqrt{z}}}+4\,{\frac{\hat{I}_{{2}}(3,2,1){{m_{b}%
}}^{5}}{\sqrt{z}}} \\
&&-4\,{\frac{\hat{I}_{2}^{[0,2]}(3,2,2){{m_{b}}}^{4}}{\sqrt{z}}}+8\,{\frac{\hat{I}_{{2}%
}(2,2,1){{m_{b}}}^{5}}{\sqrt{z}}}-{\frac{\hat{I}_{0}^{[0,1]}(3,2,2){{m_{b}}}^{4}}{%
\sqrt{z}}} \\
&&-4\,{\frac{\hat{I}_{2}^{[0,1]}(3,2,1){{m_{b}}}^{4}}{z}}+4\,{\frac{%
\hat{I}_{2}^{[0,2]}(3,2,2){{m_{b}}}^{4}}{z}}-16\,{\frac{\hat{I}_{{0}}(1,2,2){{m_{b}}}^{5}%
}{z}} \\
&&-16\,{\frac{\hat{I}_{1}^{[0,1]}(2,1,2){{m_{b}}}^{4}}{z}}-{\frac{\hat{I}_{{2}}(3,2,2){{%
m_{b}}}^{6}}{z}}+16\,{\frac{\hat{I}_{1}^{[0,2]}(3,1,2){{m_{b}}}^{3}}{z}} \\
&&-8\,{\frac{\hat{I}_{1}^{[0,1]}(3,2,1){{m_{b}}}^{4}}{z}}+3\,{\frac{\hat{I}_{{2}}(4,1,1){%
{m_{b}}}^{4}}{z}}-32\,{\frac{\hat{I}_{{2}}(2,2,1){{m_{b}}}^{5}}{z}} \\
&&+12\,{\frac{\hat{I}_{2}^{[0,1]}(3,1,2){{m_{b}}}^{4}}{z}}+16\,{\frac{%
\hat{I}_{2}^{[0,1]}(2,2,1){{m_{b}}}^{4}}{z}}-4\,{\frac{\hat{I}_{{1}}(3,2,1){{m_{b}}}^{5}%
}{z}} \\
&&-2\,{\frac{\hat{I}_{{0}}(2,2,2){{m_{b}}}^{5}}{z}}+{\frac{\hat{I}_{{0}}(3,1,2){{m_{b}}}%
^{4}}{z}}-16\,{\frac{\hat{I}_{1}^{[0,1]}(2,2,2){{m_{b}}}^{5}}{z}} \\
&&-4\,{\frac{\hat{I}_{{2}}(3,2,1){{m_{b}}}^{5}}{z}}+4\,{\frac{\hat{I}_{0}^{[0,1]}(3,1,2){%
{m_{b}}}^{3}}{z}}+2\,{\frac{\hat{I}_{0}^{[0,2]}(3,2,2){{m_{b}}}^{3}}{z}} \\
&&-64\,{\frac{\hat{I}_{{1}}(1,1,2){{m_{b}}}^{5}}{z}}-16\,{\frac{%
\hat{I}_{2}^{[0,1]}(2,2,2){{m_{b}}}^{5}}{z}}-6\,{\frac{\hat{I}_{0}^{[0,1]}(3,2,1){{m_{b}}%
}^{3}}{z}} \\
&&+4\,{\frac{\hat{I}_{1}^{[0,2]}(3,2,2){{m_{b}}}^{4}}{z}}-8\,{\frac{%
\hat{I}_{0}^{[0,1]}(2,2,1){{m_{b}}}^{3}}{z}}-16\,{\frac{\hat{I}_{2}^{[0,2]}(3,2,1){{m_{b}%
}}^{3}}{z}} \\
&&+12\,{\frac{\hat{I}_{{0}}(1,4,1){{m_{b}}}^{6}}{z}}-8\,{\frac{\hat{I}_{{0}}(2,1,2){{%
m_{b}}}^{4}}{z}}+{\frac{\hat{I}_{0}^{[0,1]}(3,2,2){{m_{b}}}^{4}}{z}} \\
&&+3/2\,{\frac{\hat{I}_{{0}}(4,1,1){{m_{b}}}^{3}}{z}}+8\,{\frac{\hat{I}_{{1}}(2,1,2){{%
m_{b}}}^{5}}{z}}+8\,{\frac{\hat{I}_{{2}}(2,3,1){{m_{b}}}^{6}}{z}} \\
&&-1/2\,{\frac{\hat{I}_{{0}}(3,2,2){{m_{b}}}^{5}}{z}}-48\,{\frac{%
\hat{I}_{2}^{[0,2]}(2,2,2){{m_{b}}}^{4}}{z}}-8\,{\frac{\hat{I}_{0}^{[0,1]}(2,2,2){{m_{b}}%
}^{4}}{z}} \\
&&+6\,{\frac{\hat{I}_{2}^{[0,1]}(4,1,1){{m_{b}}}^{3}}{z}}+64\,{\frac{%
\hat{I}_{2}^{[0,1]}(1,2,2){{m_{b}}}^{5}}{z}}-32\,{\frac{\hat{I}_{{1}}(2,1,1){{m_{b}}}^{4}%
}{z}} \\
&&+2\,{\frac{\hat{I}_{{2}}(3,1,2){{m_{b}}}^{5}}{z}}-16\,{\frac{\hat{I}_{{0}}(2,2,1){{%
m_{b}}}^{4}}{z}}-32\,{\frac{\hat{I}_{{2}}(1,2,2){{m_{b}}}^{6}}{z}} \\
&&+2\,{\frac{\hat{I}_{1}^{[0,1]}(3,2,2){{m_{b}}}^{5}}{z}}+24\,{\frac{\hat{I}_{{0}}(1,3,1)%
{{m_{b}}}^{5}}{z}}-8\,{\frac{\hat{I}_{{1}}(2,2,1){{m_{b}}}^{5}}{z}} \\
&&+12\,{\frac{(C_{{0,1}})(3,1,2){{m_{b}}}^{4}}{z}}+64\,{\frac{\hat{I}_{{2}}(1,2,1){%
{m_{b}}}^{5}}{z}}+48\,{\frac{\hat{I}_{2}^{[0,1]}(2,2,2){{m_{b}}}^{5}}{{z}^{3/2}}}
\\
&&+32\,{\frac{\hat{I}_{{1}}(2,2,1){{m_{b}}}^{5}}{{z}^{3/2}}}-8\,{\frac{\hat{I}_{{1}%
}(2,3,1){{m_{b}}}^{6}}{{z}^{3/2}}}+16\,{\frac{\hat{I}_{1}^{[0,2]}(3,2,1){{m_{b}}}%
^{3}}{{z}^{3/2}}} \\
&&-96\,{\frac{\hat{I}_{2}^{[0,1]}(1,4,1){{m_{b}}}^{6}}{{z}^{3/2}}}-32\,{\frac{\hat{I}_{{2%
}}(1,2,2){{m_{b}}}^{6}}{{z}^{3/2}}}-64\,{\frac{\hat{I}_{1}^{[0,1]}(1,2,2){{m_{b}}}%
^{5}}{{z}^{3/2}}} \\
&&+4\,{\frac{\hat{I}_{1}^{[0,1]}(3,2,1){{m_{b}}}^{4}}{{z}^{3/2}}}+16\,{\frac{%
\hat{I}_{2}^{[0,1]}(2,3,1){{m_{b}}}^{5}}{{z}^{3/2}}}+32\,{\frac{\hat{I}_{{1}}(1,2,2){{%
m_{b}}}^{6}}{{z}^{3/2}}} \\
&&+2\,{\frac{\hat{I}_{{2}}(3,2,1){{m_{b}}}^{5}}{{z}^{3/2}}}-32\,{\frac{\hat{I}_{{0}%
}(1,3,1){{m_{b}}}^{5}}{{z}^{3/2}}}+{\frac{\hat{I}_{{2}}(3,2,2){{m_{b}}}^{6}}{{z}%
^{3/2}}} \\
&&+4\,{\frac{\hat{I}_{{1}}(3,2,1){{m_{b}}}^{5}}{{z}^{3/2}}}-2\,{\frac{\hat{I}_{{2}%
}(3,1,2){{m_{b}}}^{5}}{{z}^{3/2}}}-4\,{\frac{\hat{I}_{2}^{[0,1]}(3,2,2){{m_{b}}}%
^{5}}{{z}^{3/2}}} \\
&&-4\,{\frac{\hat{I}_{1}^{[0,2]}(3,2,2){{m_{b}}}^{4}}{{z}^{3/2}}}-2\,{\frac{%
\hat{I}_{0}^{[0,1]}(3,2,2){{m_{b}}}^{4}}{{z}^{3/2}}}+12\,{\frac{%
\hat{I}_{2}^{[0,1]}(3,2,1){{m_{b}}}^{4}}{{z}^{3/2}}} \\
&&-4\,{\frac{\hat{I}_{0}^{[0,1]}(3,2,1){{m_{b}}}^{3}}{{z}^{3/2}}}-3\,{\frac{\hat{I}_{{2}%
}(4,1,1){{m_{b}}}^{4}}{{z}^{3/2}}}+16\,{\frac{\hat{I}_{1}^{[0,1]}(2,2,2){{m_{b}}}%
^{5}}{{z}^{3/2}}} \\
&&-16\,{\frac{\hat{I}_{1}^{[0,1]}(2,2,1){{m_{b}}}^{4}}{{z}^{3/2}}}+8\,{\frac{\hat{I}_{{0}%
}(2,2,1){{m_{b}}}^{4}}{{z}^{3/2}}}-12\,{\frac{\hat{I}_{2}^{[0,2]}(3,2,2){{m_{b}}}%
^{4}}{{z}^{3/2}}} \\
&&+8\,{\frac{\hat{I}_{{2}}(2,2,2){{m_{b}}}^{6}}{{z}^{3/2}}}-128\,{\frac{\hat{I}_{{2}%
}(1,3,1){{m_{b}}}^{6}}{{z}^{3/2}}}-3\,{\frac{\hat{I}_{{1}}(4,1,1){{m_{b}}}^{4}}{{z}%
^{3/2}}}\\
&&-8\,{\frac{\hat{I}_{0}^{[0,1]}(2,3,1){{m_{b}}}^{4}}{{z}^{3/2}}}-6\,{\frac{%
\hat{I}_{1}^{[0,1]}(4,1,1){{m_{b}}}^{3}}{{z}^{3/2}}}-12\,{\frac{%
\hat{I}_{1}^{[0,1]}(3,1,2){{m_{b}}}^{4}}{{z}^{3/2}}} \\
&&+4\,{\frac{\hat{I}_{{0}}(2,2,2){{m_{b}}}^{5}}{{z}^{3/2}}}-2\,{\frac{\hat{I}_{{1}%
}(3,1,2){{m_{b}}}^{5}}{{z}^{3/2}}}+48\,{\frac{\hat{I}_{1}^{[0,2]}(2,2,2){{m_{b}}}%
^{4}}{{z}^{3/2}}} \\
&&-64\,{\frac{\hat{I}_{{1}}(1,2,1){{m_{b}}}^{5}}{{z}^{3/2}}}-{\frac{\hat{I}_{{0}}(3,2,1){%
{m_{b}}}^{4}}{{z}^{3/2}}}-24\,{\frac{\hat{I}_{{0}}(1,4,1){{m_{b}}}^{6}}{{z}^{3/2}}}
\\
&&-48\,{\frac{\hat{I}_{{2}}(1,4,1){{m_{b}}}^{7}}{{z}^{3/2}}}+{\frac{\hat{I}_{{1}}(3,2,2){%
{m_{b}}}^{6}}{{z}^{3/2}}}-12\,{\frac{\hat{I}_{1}^{[0,1]}(3,2,1){{m_{b}}}^{4}}{{z}%
^{2}}} \\
&&+4\,{\frac{\hat{I}_{1}^{[0,1]}(3,2,2){{m_{b}}}^{5}}{{z}^{2}}}-16\,{\frac{%
\hat{I}_{1}^{[0,1]}(2,3,1){{m_{b}}}^{5}}{{z}^{2}}}+4\,{\frac{\hat{I}_{{0}}(2,3,1){{m_{b}}%
}^{5}}{{z}^{2}}} \\
&&+3\,{\frac{\hat{I}_{{1}}(4,1,1){{m_{b}}}^{4}}{{z}^{2}}}+128\,{\frac{\hat{I}_{{1}%
}(1,3,1){{m_{b}}}^{6}}{{z}^{2}}}-2\,{\frac{\hat{I}_{{2}}(3,2,1){{m_{b}}}^{5}}{{z}%
^{2}}} \\
&&-8\,{\frac{\hat{I}_{{1}}(2,2,2){{m_{b}}}^{6}}{{z}^{2}}}+96\,{\frac{%
\hat{I}_{1}^{[0,1]}(1,4,1){{m_{b}}}^{6}}{{z}^{2}}}-8\,{\frac{\hat{I}_{{2}}(2,3,1){{m_{b}}%
}^{6}}{{z}^{2}}} \\
&&+32\,{\frac{\hat{I}_{{1}}(1,2,2){{m_{b}}}^{6}}{{z}^{2}}}-{\frac{\hat{I}_{{1}}(3,2,2){{%
m_{b}}}^{6}}{{z}^{2}}}+2\,{\frac{\hat{I}_{{1}}(3,1,2){{m_{b}}}^{5}}{{z}^{2}}} \\
&&+12\,{\frac{\hat{I}_{1}^{[0,2]}(3,2,2){{m_{b}}}^{4}}{{z}^{2}}}+1/2\,{\frac{\hat{I}_{{0}%
}(3,2,2){{m_{b}}}^{5}}{{z}^{2}}}-2\,{\frac{\hat{I}_{{1}}(3,2,1){{m_{b}}}^{5}}{{z}%
^{2}}} \\
&&+6\,{\frac{\hat{I}_{2}^{[0,1]}(3,2,2){{m_{b}}}^{5}}{{z}^{2}}}-48\,{\frac{%
\hat{I}_{1}^{[0,1]}(2,2,2){{m_{b}}}^{5}}{{z}^{2}}}+{\frac{\hat{I}_{{2}}(3,2,2){{m_{b}}}%
^{6}}{{z}^{2}}} \\
&&-12\,{\frac{\hat{I}_{{2}}(2,2,2){{m_{b}}}^{6}}{{z}^{2}}}+48\,{\frac{\hat{I}_{{2}%
}(1,4,1){{m_{b}}}^{7}}{{z}^{2}}}+48\,{\frac{\hat{I}_{{1}}(1,4,1){{m_{b}}}^{7}}{{z}%
^{2}}} \\
&&-{\frac{\hat{I}_{{2}}(3,2,2){{m_{b}}}^{6}}{{z}^{5/2}}}-48\,{\frac{\hat{I}_{{1}}(1,4,1){%
{m_{b}}}^{7}}{{z}^{5/2}}}-{\frac{\hat{I}_{{1}}(3,2,2){{m_{b}}}^{6}}{{z}^{5/2}}} \\
&&-6\,{\frac{\hat{I}_{1}^{[0,1]}(3,2,2){{m_{b}}}^{5}}{{z}^{5/2}}}+8\,{\frac{\hat{I}_{{1}%
}(2,3,1){{m_{b}}}^{6}}{{z}^{5/2}}}+12\,{\frac{\hat{I}_{{1}}(2,2,2){{m_{b}}}^{6}}{{z%
}^{5/2}}} \\
&&+2\,{\frac{\hat{I}_{{1}}(3,2,1){{m_{b}}}^{5}}{{z}^{5/2}}}+{\frac{\hat{I}_{{1}}(3,2,2){{%
m_{b}}}^{6}}{{z}^{3}}}-2\,\hat{I}_{{0}}(3,1,1){{m_{b}}}^{3} \\
&&-\hat{I}_{{0}}(3,2,1){{m_{b}}}^{4}-4\,\hat{I}_{{0}}(2,1,2){{m_{b}}}^{4}
\end{eqnarray*}
\begin{eqnarray*}
C^{HQET}_{T} &=&32\,{\frac{\hat{I}_{{0}}(1,2,2){{m_{b}}}^{5}}{\sqrt{z}}}-4\,{\frac{%
\hat{I}_{0}^{[0,1]}(3,1,2){{m_{b}}}^{3}}{\sqrt{z}}}-4\,{\frac{\hat{I}_{0}^{[0,1]}(3,2,1){%
{m_{b}}}^{3}}{\sqrt{z}}} \\
&&+4\,{\frac{\hat{I}_{0}^{[0,2]}(3,2,2){{m_{b}}}^{3}}{\sqrt{z}}}+8\,{\frac{\hat{I}_{{0}%
}(2,3,1){{m_{b}}}^{5}}{\sqrt{z}}}-{\frac{\hat{I}_{{0}}(3,2,2){{m_{b}}}^{5}}{\sqrt{z%
}}} \\
&&+8\,{\frac{\hat{I}_{{0}}(2,2,1){{m_{b}}}^{4}}{\sqrt{z}}}-32\,{\frac{%
\hat{I}_{0}^{[0,1]}(2,1,2){{m_{b}}}^{3}}{\sqrt{z}}}+3\,{\frac{\hat{I}_{{0}}(4,1,1){{m_{b}%
}}^{3}}{\sqrt{z}}} \\
&&+8\,{\frac{\hat{I}_{0}^{[0,2]}(3,1,2){{m_{b}}}^{2}}{\sqrt{z}}}-2\,{\frac{\hat{I}_{{0}%
}(3,1,2){{m_{b}}}^{4}}{\sqrt{z}}}-16\,{\frac{\hat{I}_{0}^{[0,1]}(2,2,2){{m_{b}}}%
^{4}}{\sqrt{z}}} \\
&&-16\,{\frac{\hat{I}_{{0}}(2,1,1){{m_{b}}}^{3}}{\sqrt{z}}}+64\,{\frac{\hat{I}_{{0}%
}(1,1,2){{m_{b}}}^{4}}{\sqrt{z}}}+8\,{\frac{\hat{I}_{{0}}(2,1,2){{m_{b}}}^{4}}{%
\sqrt{z}}} \\
&&-16\,{\frac{\hat{I}_{0}^{[0,1]}(2,3,1){{m_{b}}}^{4}}{z}}-8\,{\frac{%
\hat{I}_{0}^{[0,2]}(3,2,1){{m_{b}}}^{2}}{z}}+2\,{\frac{\hat{I}_{{0}}(3,1,2){{m_{b}}}^{4}%
}{z}} \\
&&+6\,{\frac{\hat{I}_{{0}}(3,2,1){{m_{b}}}^{4}}{z}}-8\,{\frac{\hat{I}_{0}^{[0,1]}(3,1,2){%
{m_{b}}}^{3}}{z}}-48\,{\frac{\hat{I}_{{0}}(1,4,1){{m_{b}}}^{6}}{z}} \\
&&-4\,{\frac{\hat{I}_{0}^{[0,1]}(3,2,2){{m_{b}}}^{4}}{z}}+16\,{\frac{\hat{I}_{{0}}(2,1,2)%
{{m_{b}}}^{4}}{z}}+16\,{\frac{\hat{I}_{0}^{[0,1]}(3,1,1){{m_{b}}}^{2}}{z}} \\
&&+4\,{\frac{\hat{I}_{0}^{[0,1]}(3,2,1){{m_{b}}}^{3}}{z}}+8\,{\frac{\hat{I}_{{0}}(3,1,1){%
{m_{b}}}^{3}}{z}}-32\,{\frac{\hat{I}_{{0}}(2,1,1){{m_{b}}}^{3}}{z}} \\
&&-192\,{\frac{\hat{I}_{{0}}(1,2,1){{m_{b}}}^{4}}{z}}-8\,{\frac{\hat{I}_{{0}}(2,2,1){{%
m_{b}}}^{4}}{z}}+8\,{\frac{\hat{I}_{{0}}(2,2,2){{m_{b}}}^{5}}{z}} \\
&&-160\,{\frac{\hat{I}_{{0}}(1,3,1){{m_{b}}}^{5}}{z}}+2\,{\frac{\hat{I}_{{0}}(3,1,2){{%
m_{b}}}^{4}}{{z}^{3/2}}}-2\,{\frac{\hat{I}_{{0}}(3,2,1){{m_{b}}}^{4}}{{z}^{3/2}}}
\\
&&+8\,{\frac{\hat{I}_{{0}}(2,3,1){{m_{b}}}^{5}}{{z}^{3/2}}}+{\frac{\hat{I}_{{0}}(3,2,2){{%
m_{b}}}^{5}}{{z}^{3/2}}}-8\,{\frac{\hat{I}_{{0}}(3,1,1){{m_{b}}}^{3}}{{z}^{3/2}}}
\\
&&+8\,{\frac{\hat{I}_{0}^{[0,1]}(3,2,1){{m_{b}}}^{3}}{{z}^{3/2}}}-2\,{\frac{\hat{I}_{{0}%
}(3,2,1){{m_{b}}}^{4}}{{z}^{2}}}+4\,\hat{I}_{{0}}(3,2,1){{m_{b}}}^{4} \\
&&-2\,\hat{I}_{0}^{[0,1]}(3,2,2){{m_{b}}}^{4}+8\,\hat{I}_{{0}}(3,1,1){{m_{b}}}%
^{3}-12\,\hat{I}_{0}^{[0,1]}(3,1,2){{m_{b}}}^{3} \\
&&+4\,\hat{I}_{{0}}(2,2,2){{m_{b}}}^{5}+16\,\hat{I}_{{0}}(2,1,2){{m_{b}}}^{4}
\end{eqnarray*}
where
\begin{eqnarray}
\hat{I}_n^{[i,j]} (a,b,c) =\frac{(2m_b)^{i+j}}{(\sqrt{z})^{j}}
\left( T_1^2 \right)^i \left( T_2^2 \right)^j
\frac{d^i}{d\left( T_1^2 \right)^i} \frac{d^j}{d\left( T_2^2 \right)^j} %
\left[\left( T_1^2 \right)^i \left( T_2^2 \right)^j \hat{I}_n(a,b,c) \right]%
~.  \nonumber
\end{eqnarray}
\newline
\newline
\newline

\begin{figure}[th]
\vspace*{+2.5cm}
\begin{center}
\begin{picture}(140,100)
\put(0,10){ \epsfxsize=10cm \epsfbox{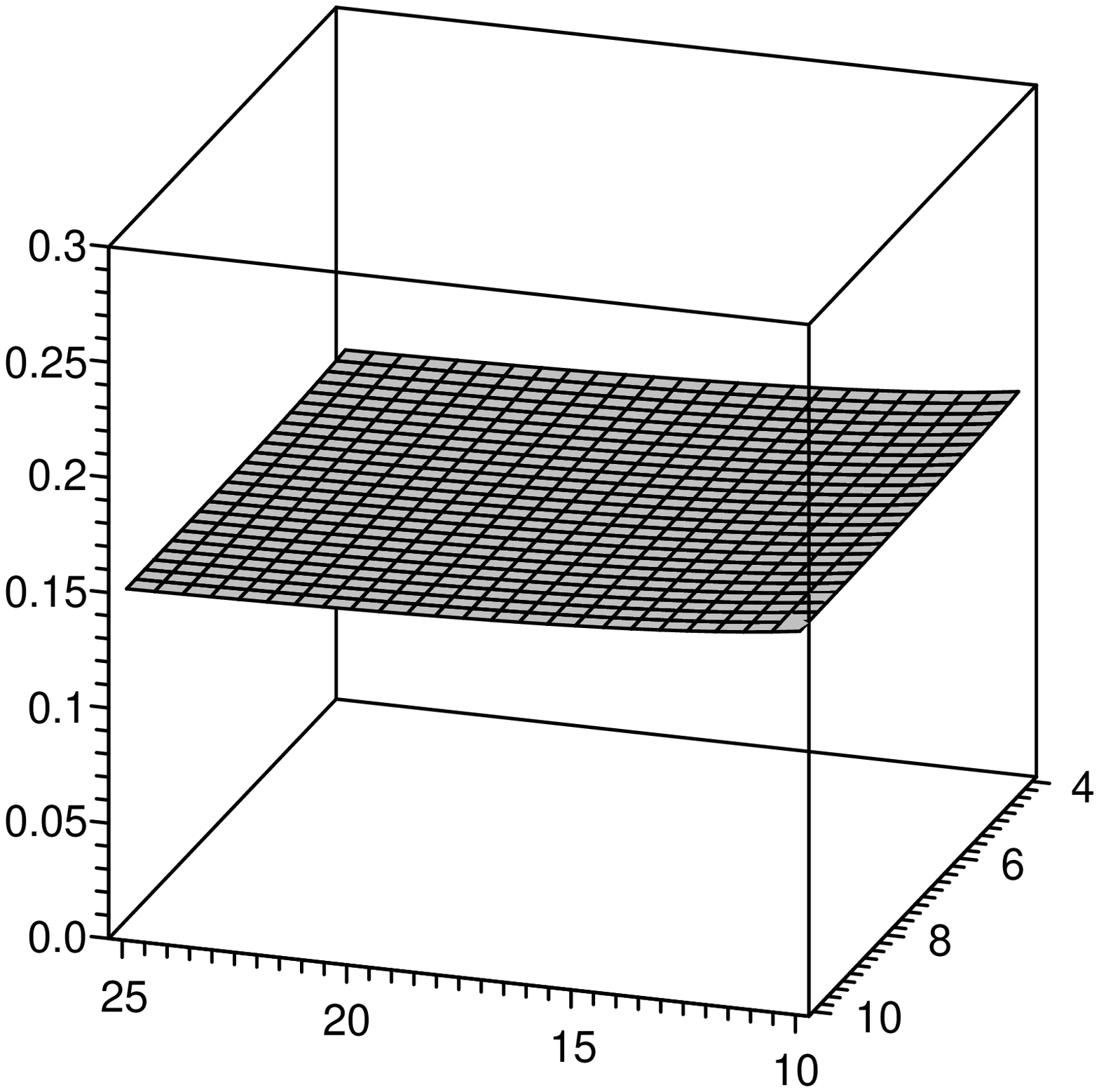}}
\put(0,95){$f_{_{+}}$} \put(80,20){$M^2_{2}$,
GeV$^2$}\put(10,20){$M^2_{1}$, GeV$^2$}
\end{picture}
\end{center}
\vspace*{-1cm} \caption{The dependence of the form factor $f_{_{+}}$
on Borel parameters $M^2_{1}$ and $M^2_{2}$ for $B_c \rar D_s
l^{+}l^{-}/\nu\bar{\nu}$.}
\end{figure}
\begin{figure}[th]
\vspace*{+2.5cm}
\begin{center}
\begin{picture}(140,100)
\put(0,10){ \epsfxsize=10cm \epsfbox{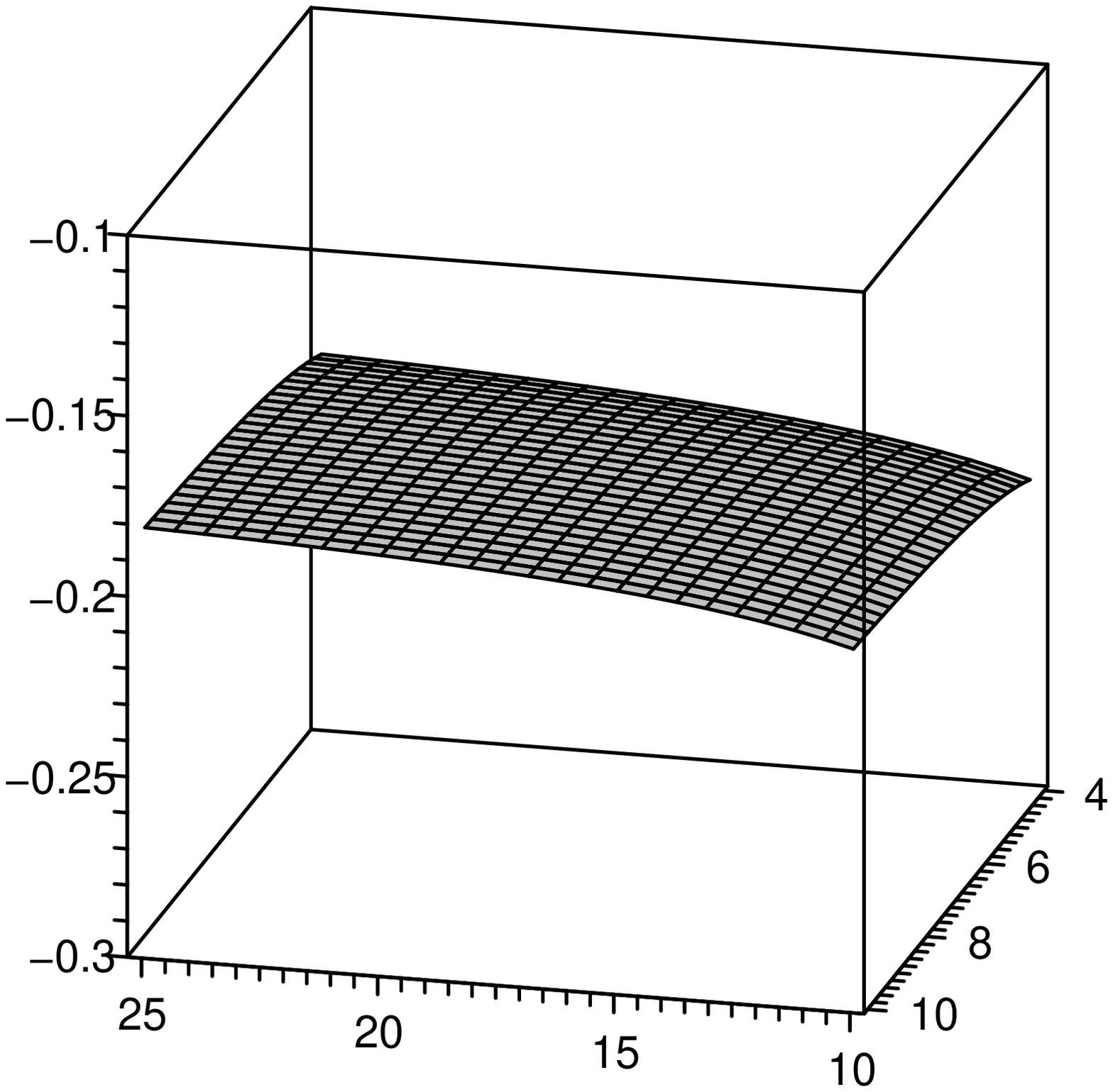}}
\put(0,95){$f_{_{-}}$} \put(80,20){$M^2_{2}$,
GeV$^2$}\put(10,20){$M^2_{1}$, GeV$^2$}
\end{picture}
\end{center}
\vspace*{-1cm} \caption{The same as Fig. 3, but for $f_{_{-}}$.}
\end{figure}
\begin{figure}[th]
\vspace*{+2.5cm}
\begin{center}
\begin{picture}(140,100)
\put(0,10){ \epsfxsize=10cm \epsfbox{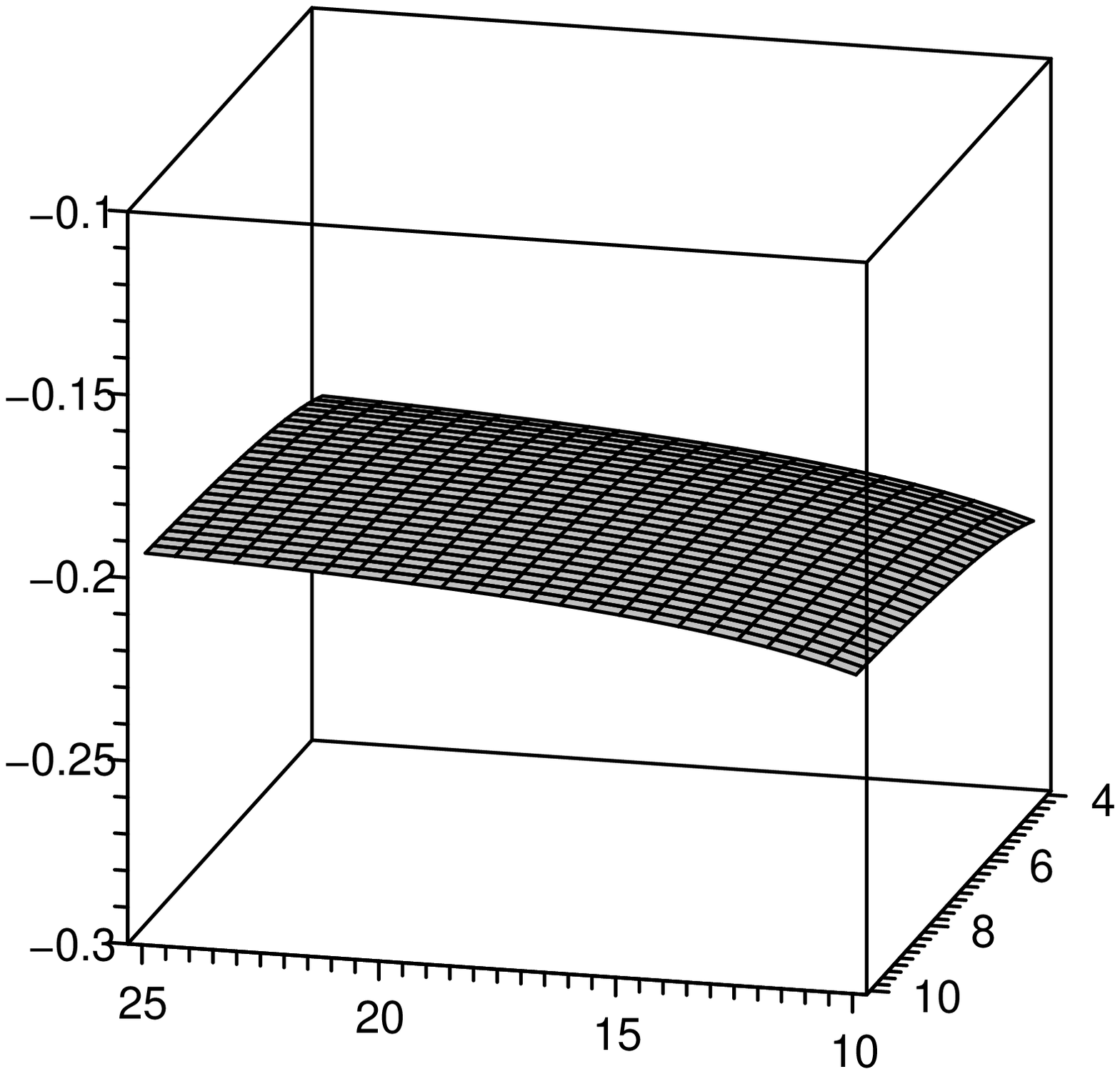}}
\put(0,95){$f_{_{T}}$} \put(80,20){$M^2_{2}$,
GeV$^2$}\put(10,20){$M^2_{1}$, GeV$^2$}
\end{picture}
\end{center}
\vspace*{-1cm} \caption{The dependence of the form factor $f_{_{T}}$
on Borel parameters $M^2_{1}$ and $M^2_{2}$ for $B_c \rar D_s
l^{+}l^{-}$.}
\end{figure}
\begin{figure}[th]
\vspace*{+2.5cm}
\begin{center}
\begin{picture}(140,100)
\put(0,10){ \epsfxsize=10cm \epsfbox{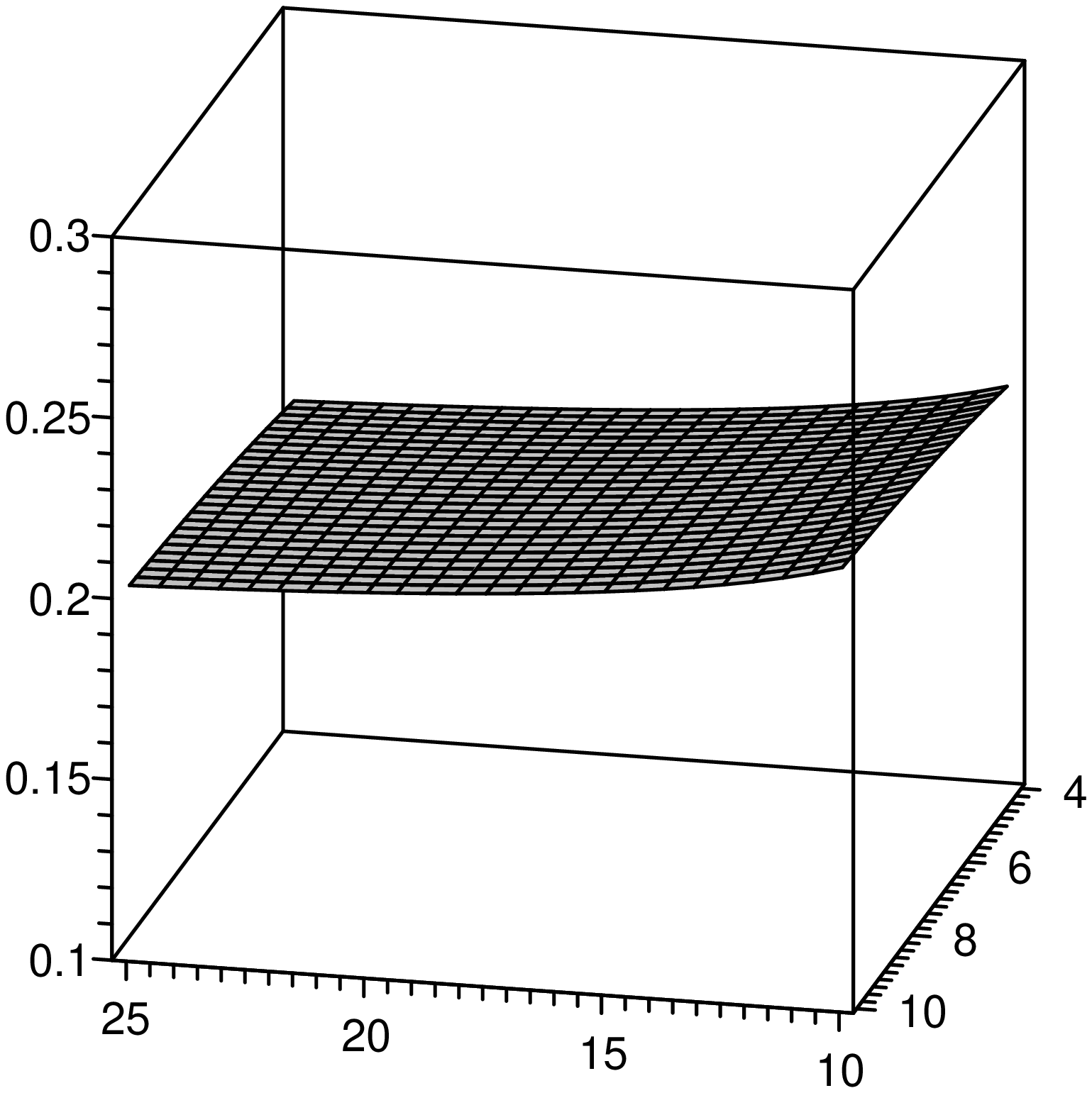}}
\put(0,95){$f_{_{+}}$} \put(80,20){$M^2_{2}$,
GeV$^2$}\put(10,20){$M^2_{1}$, GeV$^2$}
\end{picture}
\end{center}
\vspace*{-1cm} \caption{The dependence of the form factor $f_{_{+}}$
on Borel parameters $M^2_{1}$ and $M^2_{2}$ for $B_c \rar D
l^{+}l^{-}/\nu\bar{\nu}$.}
\end{figure}
\begin{figure}[th]
\vspace*{+2.5cm}
\begin{center}
\begin{picture}(140,100)
\put(0,10){ \epsfxsize=10cm \epsfbox{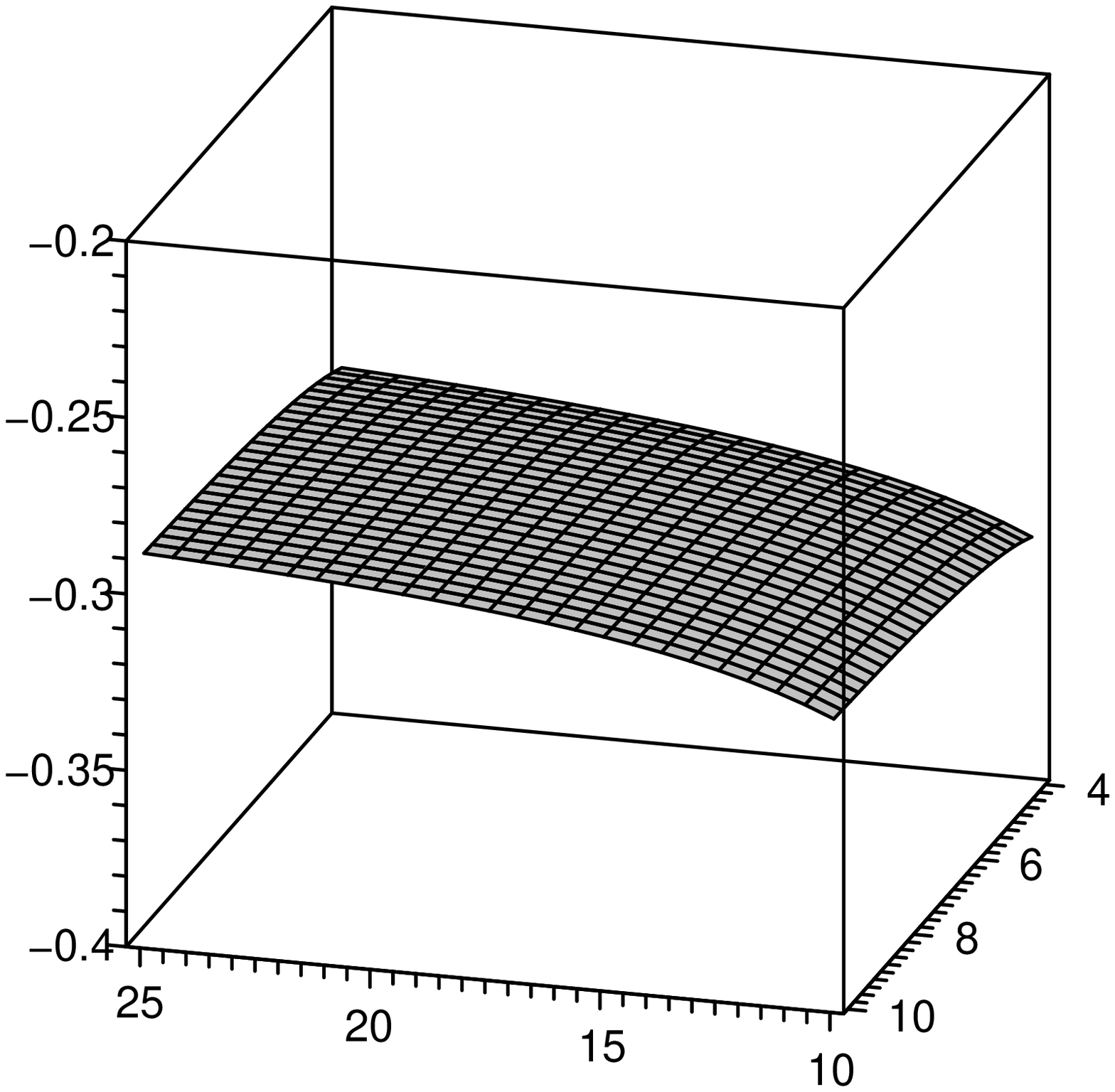}}
\put(0,95){$f_{_{-}}$} \put(80,20){$M^2_{2}$,
GeV$^2$}\put(10,20){$M^2_{1}$, GeV$^2$}
\end{picture}
\end{center}
\vspace*{-1cm} \caption{The same as Fig. 6, but for $f_{_{-}}$.}
\end{figure}
\begin{figure}[th]
\vspace*{+2.5cm}
\begin{center}
\begin{picture}(140,100)
\put(0,10){ \epsfxsize=10cm \epsfbox{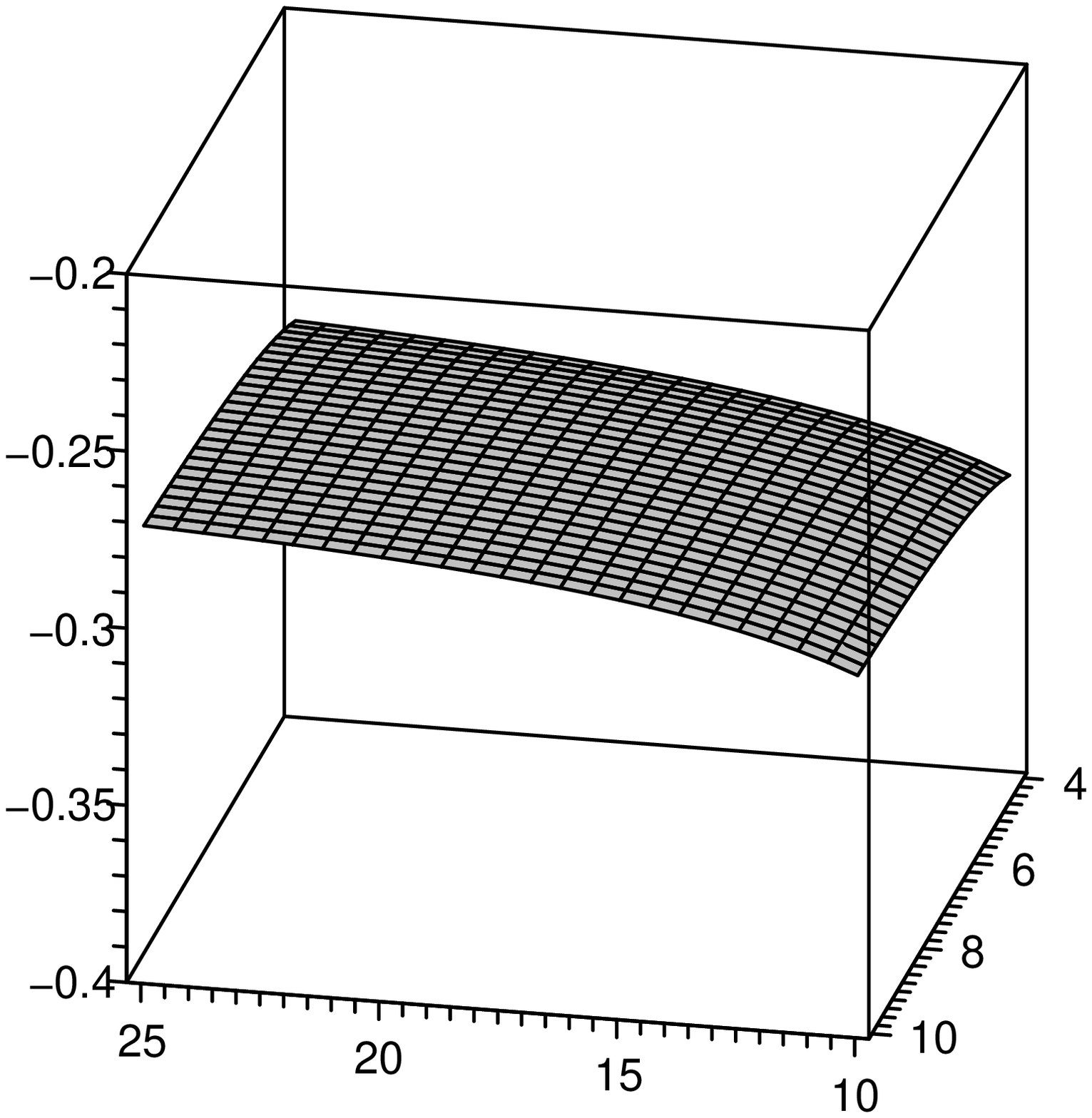}}
\put(0,95){$f_{_{T}}$} \put(80,20){$M^2_{2}$,
GeV$^2$}\put(10,20){$M^2_{1}$, GeV$^2$}
\end{picture}
\end{center}
\vspace*{-1cm} \caption{The dependence of the form factor $f_{_{T}}$
on Borel parameters $M^2_{1}$ and $M^2_{2}$ for $B_c \rar D
l^{+}l^{-}$.}
\end{figure}
\end{document}